\documentclass[manuscript]{aastex}

\def\msun{{\rm\,M_\odot}}

\newcommand{\etal}{et al.\ }

\newcommand{\lya}{Ly$\alpha$ }

\def\etal   {{et~al.}\ }

\def\et{{\it et\thinspace al.}}    

\def\zsun{{\rm\,Z_\odot}}
\def\msun{{\rm\,M_\odot}}

\def\vol#1  {{{#1}{\rm,}\ }}

\def\etal{et al.\ }

\lefthead{Cen \etal} 
\righthead{The WHIM}

\begin{document}

\title{Where Are the Baryons? II: Feedback Effects}

\author{
Renyue Cen$^{1}$ and Jeremiah P. Ostriker$^{2}$
} 
 
\footnotetext[1]{Princeton University Observatory, Princeton, NJ 08544;
 cen@astro.princeton.edu}
\footnotetext[2]{Princeton University Observatory, Princeton, NJ 08544;
 jpo@astro.princeton.edu}

\begin{abstract} 

Numerical simulations of the intergalactic medium have shown
that at the present epoch a significant fraction ($40-50\%$) of
the baryonic component should be found in the ($T\sim 10^6$K)
Warm-Hot Intergalactic Medium (WHIM) - with 
several recent observational lines of evidence indicating the validity 
of the prediction.
We here recompute the evolution of the WHIM
with the following major improvements:
(1) galactic superwind feedback processes from galaxy/star formation 
are explicitly included;
(2) major metal species (O V to O IX)
are computed explicitly in a non-Equilibrium way;
(3) mass and spatial dynamic ranges are larger by
a factor of 8 and 2, respectively, than in our previous simulations.
Here are the major findings:
(1) galactic superwinds have dramatic effects, increasing
the WHIM mass fraction by about $20\%$, primarily through
heating up warm gas near galaxies 
with density $10^{1.5}-10^4$ times the mean density.
(2) the fraction of baryons in WHIM 
is increased modestly from the earlier work but is $\sim 40-50\%$.
(3) the gas density of the WHIM is broadly peaked at a density $10-20$ times 
the mean density, ranging from underdense 
regions to regions that are overdense by $10^3-10^4$.
(4) the median metallicity of the WHIM 
is $0.18\zsun$ for oxygen with 50\% and $90\%$ intervals
being (0.040,0.38) and (0.0017,0.83).

\end{abstract}
 
 
\section{Introduction}

Cosmological hydrodynamic simulations have strongly suggested
that most of the previously ``missing" baryons may be in 
a gaseous phase in the temperature range
$10^5-10^7$K and at moderate overdensity 
(Cen \& Ostriker 1999, hereafter "CO"; Dav\'e 2001),
called the warm-hot intergalactic medium (WHIM),
with the primary heating process being
hydrodynamic shocks from the formation 
of large-scale structure at scales
currently becoming nonlinear.
The reality of the WHIM has now been quite convincingly
confirmed by a number of observations 
from HST, FUSE, Chandra and XMM-Newton
(Tripp, Savage, \& Jenkins 2000;
Tripp \& Savage 2000;
Oegerle \etal 2000;
Scharf \etal 2000;
Tittley \& Henriksen 2001;
Savage \etal 2002;
Fang \etal 2002; Nicastro \etal 2002;
Mathur, Weinberg, \& Chen 2002;
Kaastra \etal 2003;
Finoguenov, Briel, \& Henry 2003;
Sembach \etal 2004; Nicastro \etal 2005).

In addition to shock heating, feedback processes following
star formation in galaxies can heat gas to the same WHIM temperature range.
What is lacking theoretically is 
a satisfactory understanding of the known feedback processes
on the WHIM and how the WHIM may be used to 
understand and calibrate the feedback processes.
Another unsettled issue is how the predicted results
will change if one has a more accurate,
non-equilibrium calculation of the major
metal species, such as O VI, O VII, O VIII,
since time scales for ionization and recombination 
are not widely separated from the Hubble time scale.
We have investigated these important processes.
Additional improvements include
significantly larger dynamic ranges
of the simulation, a WMAP normalized cosmological model
and an improved radiative transfer treatment.
Our current work significantly extends previous theoretical works
by our group and others 
(CO; Dav\'e 2001;
Cen \etal 2001; Fang \etal 2002; 
Chen \etal 2003;
Furlanetto \etal 2004,2005a,b;
Yoshikawa \etal 2003; Ohashi \etal 2004; Suto \etal 2004; Fang \et
al. 2005).
In this paper we focus on the feedback effects due to star formation
and in a companion paper (Cen \& Fang 2005) we will present
additional effects due to non-equilibrium treatments of metal species
and observables of the WHIM.
The outline of this paper is as follows:
the simulation details are given in \S 2;
in \S 3 we give detailed results
and discussion and conclusions are presented in \S 4.

\section{Simulations}\label{sec: sims}

Numerical methods of the cosmological hydrodynamic code
and input physical ingredients
have been described in detail in an earlier paper
(Cen \etal 2003).
Briefly, the simulation integrates five sets of equations
simultaneously:
the Euler equations for gas dynamics in comoving coordinates,
time dependent rate equations for different hydrogen and helium
species at different ionization states,
the Newtonian equations of motion for dynamics of collisionless
(dark matter) particles,
the Poisson equation for obtaining the gravitational potential field and
the equation governing the evolution of the intergalactic
ionizing radiation field,
all in cosmological comoving coordinates.
Note that the cosmological (frequency dependent)
radiation field is solved for self-consistently, rather than
being a separate input to the modeling.
The gasdynamical equations are solved using
the TVD (Total Variation Diminishing) shock capturing code 
(Ryu \etal 1993) on an uniform mesh.
HI, HeI, HeII are separately followed as different species in a
non-LTE rate equations for each cell at every timestep.
Oxygen in various states of ionization 
are also followed as separate species (see Appendix A)
and, in total, seven species are followed in non-LTE way.
The rate equations are treated using sub-cycles within a hydrodynamic time
step due to the much shorter ionization time-scales
(i.e., the rate equations are very ``stiff").
Dark matter particles are advanced in time using the standard
particle-mesh (PM) scheme.
A leapfrog scheme integration scheme is used.
The gravitational potential on an uniform mesh is solved
using the Fast Fourier Transform (FFT) method.

The initial conditions adopted are those
for Gaussian processes with the phases
of the different waves being random and uncorrelated.
The initial condition is generated by the
COSMICS software package kindly provided by E. Bertschinger (2001).

Cooling and heating
processes due to all the principal line and continuum atomic
processes for a plasma of primordial composition with
additional metals ejected from star formation (see below),
Compton cooling due to the microwave background
radiation field and Compton cooling/heating due to
the X-ray and high energy background
are computed  in a time-dependent, non-equilibrium fashion.
The cooling due to metals is computed
using a code based on the Raymond-Smith code (Raymond, Cox, \& Smith 1976)
assuming ionization
equilibrium (Cen \etal 1995).

We follow star formation using
a well defined  
prescription
used by us in our earlier work
(Cen \& Ostriker 1992,1993) and
similar to that of other investigators
(Katz, Hernquist, \& Weinberg 1992;
Katz, Weinberg, \& Hernquist 1996;
Steinmetz 1996;
Gnedin \& Ostriker 1997).
A stellar particle of mass
$m_{*}=c_{*} m_{\rm gas} \Delta t/t_{*}$ is created
(the same amount is removed from the gas mass in the cell),
if the gas in a cell at any time meets
the following three conditions simultaneously:
(i) contracting flow, (ii) cooling time less than dynamic time, and  (iii)
Jeans unstable,
where $\Delta t$ is the time step, $t_{*}={\rm max}(t_{\rm dyn}, 10^7$yrs),
$t_{dyn}=\sqrt{3\pi/(32G\rho_{tot})}$ is the dynamical time of the cell,
$m_{\rm gas}$ is the baryonic gas mass in the cell and
$c_*=0.07$ is star formation efficiency.
Each stellar particle is given a number of other attributes at birth, including
formation time $t_i$, initial gas metallicity
and the free-fall time in the birth cell $t_{dyn}$.
The typical mass of a stellar particle in the simulation
is about one million solar masses;
in other words, these stellar particles are like
coeval globular clusters,
whose evolution can be computed with standard stellar evolution
codes such as those of Bruzual \& Charlot (1993).
Changing the numerical coefficient $c_*$ to higher or lower
values can change the time dependence of star formation, making
it more smooth or more ``bursty" but has little effect on
the total mass in stars/galaxies or in the overall simulation. 
All variations of this commonly adopted 
star-formation algorithm essentially achieve the same goal:
in any region where gas density exceeds the stellar density, gas
is transformed to stars on a timescale longer
than the local dynamical time
and shorter than the Hubble time. 
Since these two time scales are widely separated,
the effects,
on the longer time scale,
of changing the dimensionless numbers 
(here $c_*$)
are minimal.
Since nature 
does not provide us with examples
of systems which violate
this condition (systems which persist over many dynamical and cooling 
time scales  in having more gas than stars),
this commonly adopted algorithm should be 
adequate even though our detailed understanding of star formation remains primitive.

Stellar particles are treated dynamically
as collisionless particles subsequent to their birth.
But feedback from star formation is allowed in three forms:
ionizing UV photons, the effects of the cumulative 
SN explosions and AGN output known as Galactic Superwinds (GSW),
and metal-enriched
gas, all being proportional to the local star formation rate.
The temporal release of all three feedback components at time $t$
has the same form:
$f(t,t_i,t_{dyn}) \equiv (1/ t_{dyn})
[(t-t_i)/t_{dyn}]\exp[-(t-t_i)/t_{dyn}]$. Within a time step $dt$, the
released GSW energy to the IGM, ejected mass from stars into the IGM and
escaping UV radiation energy are
$e_{GSW} f(t,t_i,t_{dyn}) m_* c^2 dt$,
$e_{mass} f(t,t_i,t_{dyn}) m_* dt$
and
$f_{esc}(Z) e_{UV}(Z) f(t,t_i,t_{dyn}) m_* c^2 dt$.
We use the Bruzual-Charlot population synthesis code 
(Bruzual \& Charlot 1993; Bruzual 2000)  
to compute the intrinsic metallicity-dependent
UV spectra from stars with Salpeter IMF (with a lower and upper mass 
cutoff of $0.1\msun$ and $125\msun$).
Note that $e_{UV}$ is no longer just a simple normal coefficient but
a function of metallicity.
The Bruzual-Charlot code gives
$e_{UV}=(1.2\times 10^{-4}, 9.7\times 10^{-5}, 8.2\times 10^{-5}, 7.0\times 10^{-5}, 5.6\times 10^{-5}, 3.9\times 10^{-5} ,1.6\times 10^{-6})$ at
$Z/\zsun=(5.0\times 10^{-3}, 2.0\times 10^{-2}, 2.0\times 10^{-1}, 4.0\times 10^{-1}, 1.0, 2.5 ,5.0)$. 
We also implement a gas metallicity dependent ionizing photon
escape fraction from galaxies in the sense that higher metallicity, hence
higher dust content, galaxies are assumed to allow a lower
escape fraction; we adopt the escape fractions of
$f_{esc}=2\%$ and $5\%$ (Hurwitz \etal 1997; Deharveng \etal 2001;
Heckman \etal 2001; Steidel \etal 2001; Shapley \etal 2005)
for solar and one tenth of solar metallicity, respectively,
and interpolate/extrapolate using a linear log form of metallicity. In
addition, we include the emission from
quasars using the spectral form observationally
derived by Sazonov, Ostriker, \& Sunyaev (2004),
with a radiative efficiency in terms of stellar mass
of $e_{QSO}=2.5\times 10^{-5}$ for $h\nu>13.6$eV.
This number, $e_{QSO}$, convolves together 
the ratio of black hole mass to stellar mass,
$\sim 1\times 10^{-3}$ (Kormendy \& Gebhardt 2001),
the radiative efficiency of
black holes $0.1$ (Yu \& Tremaine 2002)
and the fraction of the radiative energy emitted
beyond the Lyman limit $0.25$
(Sazonov \etal 2004).
Finally, hot, shocked regions (like clusters of galaxies)
emit ionizing photons due to bremsstrahlung radiation,
which are also included.
The UV component is simply averaged over the box,
since the light propagation time across our box
is small compared to the time steps.
The radiation field (from $1$eV to $100$keV)
is followed in detail with allowance for
self-consistently produced radiation sources and sinks in the simulation box
and for cosmological effects, i.e., radiation transfer
for the mean field $J_\nu$ is computed with stellar,
quasar and bremsstrahlung sources and sinks due to \lya clouds etc
(Equation 7 of Cen 1992).
In addition, a local optical depth approximation is adopted to crudely mimic
the local shielding effects: each cubic cell is flagged with six
hydrogen ``optical depths" on the six faces, each equal to the product of
neutral hydrogen density, hydrogen ionization cross section and scale
height,
and the appropriate mean from the six values is then calculated;
equivalent ones for neutral helium and singly-ionized helium are also
computed. In computing the global sink terms for the radiation field
the contribution of each cell is subject to the shielding
due to its own ``optical depth".
In addition, in computing the local ionization and cooling/heating balance
for each cell the same shielding is taken into account
to attenuate the external ionizing radiation field.

It is very difficult to accurately model the effects of GSW.
A proper treatment would include allowance for a multiphase
medium having most of the mass in clouds or filaments occupying 
a small fraction of the volume.
Significant progress has been made recently
to provide a better treatment of the multi-phase interstellar
medium (Yepes \etal 1997;
Elizondo \etal 1999a,b;
Hultman \& Pharasyn 1999;
Ritchie \& Thomas 2001;
Springel \& Hernquist 2003) but
the generation of GSW is far from being adequately modeled.
Clearly, a combination of both high resolution and
detailed multi-phase medium treatment
(including magnetic fields and cosmic rays)
is requisite before our understanding of the interactions
between galaxy formation and IGM
can be considered to be truly satisfactory.
This problem is too difficult for us to address with our 
code, so we have chosen to not attempt to calculate
the {\it causes} and generation of GSW, but,
instead,
to simply assume an input level of mass,
energy and metals, and carefully compute
the {\it consequences} of GSW on the surrounding medium.
For this purpose our code 
is very well designed.

Thus, GSW energy and ejected metals
are distributed into 27 local gas cells centered at the stellar particle in
question, weighted by the specific volume of each cell.
We fix $e_{mass}=0.25$,
i.e., 25\% of the stellar mass is recycled.
GSW energy injected into the IGM is included with an adjustable
efficiency (in terms of rest-mass energy of total formed stars) of
$e_{GSW}$, which is normalized to observations for our fiducial simulation
with $e_{GSW}=3\times 10^{-6}$.
If the ejected mass and associated energy propagate into a vacuum, the
resulting velocity of the ejecta would be
$(e_{GSW}/e_{mass})^{1/2}c=1470$km/s.
After the ejecta has accumulated an amount of mass comparable to its initial
mass, the velocity may slow down to a few hundred km/s.
We find that this velocity roughly corresponds to the observed
outflow velocities of LBGs (e.g., Pettini \etal 2002).
We also make simulations with no GSW and with stronger GSW
to investigate the effects of GSW on IGM.
Thus all the uncertainties concerning how much of the energy
which is generated by feedback can escape into the surrounding
IGM is encapsulated into one parameter $e_{GSW}$.

The results reported on here are based on new simulations of
a {\it WMAP}-normalized (Spergel \etal 2003; Tegmark \etal 2004)
cold dark matter model with a cosmological constant, 
$\Omega_M=0.31$, $\Omega_b=0.048$, $\Omega_{\Lambda}=0.69$, $\sigma_8=0.89$,
$H_0=100 h {\rm km s}^{-1} {\rm Mpc}^{-1} = 69 {\rm km} s^{-1} {\rm Mpc}^{-1}$ 
and $n=0.97$.
The adopted box size is $85$Mpc/h comoving and with a number of cells
of $1024^3$, the cell size is $83$kpc/h comoving,
with dark matter particle mass equal to $3.9\times 10^8h^{-1}\msun$.
Given a lower bound of the temperature for almost
all the gas in the simulation ($T\sim 10^4$~K),
the Jeans mass $\sim 10^{10}\msun$ for mean density gas,
which is comfortably larger than our mass resolution.

\section{Results}

As in CO, in our analysis
we divide the IGM into three components by temperature:
(1) $T > 10^7$K (the normal X-ray emitting gas, predominantly
in collapsed and virialized clusters of galaxies);
(2) $10^7~$K$>T>10^5~$K gas, 
which is defined as WHIM and exists mainly in unvirialized regions;
(3) $T<10^5~$K warm gas, which is mostly in low
density regions.
A last component (4) is the cold gas that has been condensed
into stellar objects, which we designate ``galaxies",
and which will contain stars and cold gas.
After reionization at $z\sim 6$ (e.g., Gnedin \& Ostriker 1997;
Fan \etal 2002; Cen \& McDonald 2002),
driven by the absorption of UV photons from the early generation of stars and quasars,
the photoionized IGM is left in the ``warm" component with 
$T\sim 10^{4.0-4.3}$~K.
Then, as larger and larger scale structure forms,
the breaking waves of wavelength $\lambda_{NL}(t)$ cause
gas to shock heat to temperatures
${kT(t)\over \mu m_p}=A[\lambda_{NL}(t)H(t)]^2$,
which increases monotonically with time,
where $H(t)$ is the Hubble constant at time $t$.
Details of this fit were provided in CO ($A\sim 3/16$).
Thus, warm gas is increasingly shock heated,
removed from that category and
added to the WHIM category with the shocks forming the observed
filamentary network seen in Figure 2.
By redshift $z\sim 2$, $\lambda_{NL}$ has 
increased to the point where, at the nodes, where filaments intersect,
gas has been shocked into the 
``hot" X-ray emitting gas category  as we begin to see the formation
of the great clusters of galaxies in these regions.
While not expected to be the primary heater of IGM,
GSW from star forming galaxies are observed to blow winds
at speeds of several hundred kilometers per second
and will heat up IGM, in the vicinity of galaxies.

Figure 1 shows the evolution of these four components.
We note that $49\%$ of all baryons at
are in WHIM by the time $z=0$,
which is consistent with our previous findings (CO).
Without galactic superwinds the mass fraction in WHIM
is reduced from $\sim 50\%$ to $\sim 40\%$,
which may partially explain lower WHIM mass fractions
in other simulations without galactic superwinds
as found in Dav\'e \etal (2001).
The bottom line is this:
while galactic superwinds are subdominant to
gravitational heating caused by the collapse of large-scale density waves,
they make, nevertheless,
about a $20\%$ contribution to the overall WHIM mass.
Other components are consistent with our earlier results
and approximately in agreement with observations
with regard to low-z redshift \lya cloud gas (Penton, Stocke, \& Shull 2004),
X-ray clusters of galaxies and stellar mass (Fukugita, Hogan, \& Peebles 1998).

The hot component gas reached 11\% in this simulation by $z=0$,
which is somewhat smaller than the $19\%$ found in our previous work (CO).
The difference is partly due to difference in the model parameters
and partly due to cosmic variance.
This component is not greatly affected by GSW (cf. Figure 13c).
As noted, our attention will be focused in this paper 
on the WHIM (solid dots) in Figure 1b:
the warm/hot gas rises dramatically in abundance
with increasing time and dominates
the mass balance by $z=0$,
reaching 49\% of the total baryons.
Therefore, our new simulations with a host of improved
physical treatments and numerical details
confirm our previous results:
the long sought missing baryons (those not seen in the more easily
observed warm and hot components) should be found in the WHIM.
The primary purpose of this paper is, in addition, 
to provide a more quantitative description
of the observable properties of the WHIM to better test this prediction.

All subsequent results are focused on redshift zero.
The characteristic spatial distribution 
of the WHIM is shown in Figure 2.
We see again  a filamentary structure (in green) as found previously (CO)
where, at the high density nodes (red), 
``galaxies" as well as very hot gas have been collected to form 
X-ray bright clusters of galaxies.
This ``Cosmic Web" is traced out by moderate density
gas (the green filaments); 
galaxies and other virialized objects are beads 
threading the green filaments.
The vast volume between the filaments 
is mostly underdense in gas, which is also
relatively devoid of galaxies (Peebles 1999).

Figure 3 shows the gas distribution in 
the density-temperature plane. 
The gas in this phase space can be best described using
the four baryon components defined in the beginning of this section.
First, one sees a cooling feature at $T\sim 10^4$~K and overdensity $\sim 10^{3.5-5.0}$,
where the gas is cooling rapidly to form stars.
Second, the lower left corner in the plot 
($\rho \le 1$ and $T<10^4$~K) represents cold-warm gas in the
voids which does not cool efficiently via radiative processes
but cools adiabatically due to the expansion of the universe 
after initial photoionization and photoheating.
Third, both WHIM and hot gas are heated up primarily by
shocks formed during large-scale structure formation.
Initially, this heated gas tends to occupy the upper right
corner of the phase space, representing relatively dense regions
that have attained high temperature due to shocks,
some of which are in virialized regions corresponding to 
groups and clusters of galaxies.
With time the shocks continue to expand and propagate into
lower density regions when matter continues to infall towards
these regions, resulting in the mass concentration contours in Figure 3
moving to the left 
in the displayed phase space and filling up
the upper left quarter of the plot.
We provide a fitting formula that traces the locus of the ridge line 
in the $T-\rho$ plane for the WHIM as:
\begin{equation}
\log(T) = 8 - [4(\log\rho_g/\bar\rho_g)^{0.9}]^{-1},
\end{equation}
\noindent
where $T$ is in the range $10^5-10^7$~K 
and $\rho_g$ is gas density
and $\bar\rho_g$ is the mean gas density $z=0$.

Comparing the top and bottom panels in Figure 3,
we see that, while the overall effect of GSW
on the IGM is significant as shown in Figure 1b,
it is not easily discernible here quantitatively.
However, one feature is very noticeable even to the eye:
it is clearly seen that in the simulation with 
galactic superwinds some gas in IGM  
is being pushed towards the left (near the upper left corner)
in the WHIM temperature range in the plot.
This suggests that GSW
visibly propagate into lower density regions.

Before presenting detailed quantitative calculations of the GSW effects,
we provide a more pictorial view of the GSW processes.
Figures 4-9 provide a close-up 
view of a region of size $21.2\times 21.2$Mpc$^2$/h$^2$
and thickness of $1.75$Mpc/h that has had significant GSW activity
in the recent past. 
We note that, as shown in Cen, Nagamine \& Ostriker (2005),
GSW effects tend to be more vigorous at higher redshifts
when star formation rate was higher.
Nevertheless, the GSW effects on the surrounding IGM 
are strongly visible even at $z=0$.
Figure 4 compares the gas density distributions
in the two simulations without and with GSW, respectively.
Overall, the effect on the appearance of large-scale density structure 
by GSW is small 
and its effect on low density regions (blue and purple regions) is negligible,
simply because GSW do not reach 
there or, they turn very weak even if reaching there.
The effect on high density regions ($>10^2$; yellow regions)
is quite visible, in the sense that GSW tend to suppress the gas density
concentration in these regions and disperse gas outwards,
noting that some of the yellow spots in the simulation without GSW
are significantly suppressed;
examples include the features at $(8,5)$Mpc/h, $(13,8)$Mpc/h and $(6,13)$Mpc/h.
This effect will be quantified more precisely in subsequent figures.

Figure 5 shows the temperature distribution in the same slice
for simulations without and with GSW, respectively.
We see that large-scale gravitational collapse induced 
shocks tend to center on dense regions;
these are virialization and infall shocks due to gravitational collapse 
of high density peaks.
Some of the larger peaks are seen to be enclosed by shocks of temperature
in excess of $10^7$~K (note that the displayed picture is inevitably
subject to smoothing by projection thus the higher temperature regions 
have their temperatures somewhat underestimated).
A few shock structures are clearly caused by GSW, however,
because they appear prominent only in the simulation with GSW;
for example, in the upper left quarter,
the NW-SE oriented gravitational shock feature near $(6,14)$Mpc/h, 
which is aligned with a high density filament (see Figure 4),
is significantly enlarged by GSW shocks 
propagating approximately in the direction (NE-SW) perpendicular 
to the filament.
The resulting shock temperature is in the WHIM temperature range
of $10^5-10^7$~K.
Another large feature is seen in the lower right quarter 
near $(18,5)$Mpc/h in the simulation with GSW,
which is nearly invisible in the simulation without GSW;
this feature is most likely produced by a GSW shock originating
from a galaxy not centrally located in the displayed slice,
since there is no significant density concentration at
that location in Figure 4.
These regions affected by GSW clearly are heating up regions of warm gas 
and its surroundings seen in Figures 4,5 and adding this gas to 
the WHIM; more quantitative results are given later.

Figure 6 shows the metals density distribution in the same slice
for simulations without and with GSW, respectively.
The most visible difference between simulations with and without GSW
is that the high metallicity (yellow) regions in the close vicinity of galaxies 
are substantially extended by GSW.
This is because metals in the case of no GSW are
concentrated in regions of about our grid size
and the GSW cause these extremely dense metal nuggets to appear on the IGM scale.
The predicted effect has perhaps received 
dramatic confirmation in the O VI absorption measurements
of Stocke \etal (2005),
where they find that O VI absorbers 
are invariably found with $800$~kpc of galaxies.
The affected regions have a size of a few hundred kiloparsecs to about 
one megaparsec, suggesting that this is the range of influence of GSW
in transporting metals to the IGM.
It is noted that the shocks, seen in Figure 5,
tend to propagate further out to scales of $1-2$Mpc/h,
while apparently leaving metals significantly behind,
i.e., the energy of the GSW is as expected
transported more efficiently than metals.
As we will show quantitatively later,
heating metal-rich warm gas and significant dispersal of heated-up
WHIM gas to significant distances from galaxies 
is the most dramatic role played by GSW.
Most spectacularly, GSW reduces the total amount of metals in
cold-warm gas in the close vicinity of galaxies 
by a factor of $3$ compared to the case without GSW
and adds all this metal-enriched gas to WHIM.

Figure 7 compares the non-LTE-computed  O VI density 
for the simulations without and with GSW, respectively.
The visibly discernible difference caused by GSW is
that GSW tend to disperse concentrations of O VI density,
in accordance with overall metals dispersal in dense regions seen in 
Figures 6, and tend to create typically non-spherical features
in the immediate regions surrounding galaxies on the scales of $<1$Mpc.
The differences caused by GSW on O VII density shown
in Figure 8 
appear to be somewhat larger than those for O VI density,
suggesting that GSW heated regions around galaxies
are more often in the range of $10^6-10^7$~K,
higher than optimal temperature of $10^{5.5}$~K for O VI,
which seems to be consistent with Figures 5.
This is confirmed by Figure 9 
where O VIII density is shown for simulations without and with GSW, respectively.
The overall results from Figures 4-9 indicate
that the effects of GSW on the abundances of individual 
species can not be easily described and the only means to compare 
models and observations is via direct simulations.
This is because the relative abundance of an individual 
species depends on the temperature and density history of the gas, which
in turn would depend on the shock (i.e., GSW) strengths and
environments, which are too complicated to be characterized by
simple analytically tractable relations.

Clearly, the regions where differences caused by GSW 
are visible to the eye in Figure 3
do not contain a significant amount of IGM mass.
More dense regions must have been affected by 
galactic superwinds to account for the 
effect indicated in Figure 1b,
which are visually confirmed by the pictures shown above.
We quantify this further to have a better understanding.
Figures 10,11,12 show
the concentration of gas mass of the three IGM components,
warm, WHIM and hot, respectively, in the dark matter density-gas density plane.
Comparing the simulations with (solid contours) and without (dashed contours)
GSW reveals some important information.
From Figure 10 we see that there are two separate mass concentrations
for the warm IGM component, one centered 
at about the mean density and the other at about $10^3$ times the mean density.
The physical nature of these two separate concentration is quite clear, as noted ealier:
the peak at the lower density is primarily photo-heated
gas in the voids, whereas the peak at the high density is the cooled 
gas reservoir for star formation.
The effect of GSW seems to push the peak at the lower density
to the left, i.e., some gas in low-to-moderate density
is being pushed to slightly lower (dark matter) density regions by 
GSW. This is quite understandable.
The peak at the high density, however,
appears to move both left and down-left with GSW.
This may be understood by two cases that may be occuring simultaneously. 
First, as in the previous case, gas is simply 
blown out of high density regions, causing it to move left.
Second, for some regions where gas density has become comparable to
or larger than the dark matter density due to gas cooling and condensation,
reduction in gas density
due to GSW would have dynamic effects on both the remaining gas
and the dark matter, 
resulting in a gravitational potential change, which may
explain the down-left movement of the contour peak.
Since the high density regions provide the gas reservoir
for star formation, GSW is explicitly shown here
to have a major effect on reducing and regulating star formation,
which is verified by the reduced (but adequate) star formation in the simulation
with GSW that appears to be  consistent with observations (Fukugita, Hogan \& Peebles 1998);
the total stellar mass density at $z=0$ is $\Omega_*=0.0037$,
as compared to $0.006$ in the simulation with GSW.

The effect of GSW on the WHIM is shown in Figure 11. 
Since most of the mass in WHIM is in regions with moderate 
density ($10-20$ times the mean density),
the dark matter dominates the  gravitational potential.
Therefore, the primary effect of the GSW on WHIM is in moving 
gas into somewhat lower density regions and heating up gas in 
regions with a wide range of densities, as shocks sweep through them 
in the direction of negative density gradient.
The same effect is seen for the hot component (Figure 12),
although the overall effect is somewhat diminished,
simply because the incremental increase in energy due to
GSW in these regions is subdominant to the gravitational potential energy,
released in strong shocks.

Figure 13 summarizes more quantitatively the trends in
Figures 10,11,12, in a slightly different way,
where the probability distribution of each of the three
gas components is shown as a function of the gas density.
The most dramatic difference is seen for the warm component (dotted curves in the top panel)
between simulations with (thick dotted curve)
and without (thin dotted curve) GSW.
We see that a substantial fraction of the warm gas at density 
$\rho/<\rho_b>=10^{1.3}-10^{4.5}$ is removed by GSW.
This removed gas from the warm component is primarily added to the 
WHIM gas, as clearly indicated by the difference
for WHIM components between 
simulations with (thick solid curve)
and without (thin solid curve) GSW.
And there is an overall shift of this component to lower densities
where radiative cooling would be less efficient.
The effect of GSW on the hot component is rather small,
as seen by the nearly overlapping thick dashed curve
and thin dashed curve in the bottom panel.
Figure 14 demonstrates in a different way the same point:
the gain in WHIM fraction is at the expense of warm gas,
heated up by GSW.

Metals originated in galaxies
are transported by GSW and other hydrodynamic/gravitational processes.
The role that GSW actually play to affect the distribution of metals in the IGM
are complex.
Figure 15 shows the distribution of metals in each of the three IGM components
as a function of metallicity.
The warm component displays a bimodal distribution with one peak 
at high metallicity corresponding to the star formation regions,
while the other peak at very low metallicity corresponding to
the uncontaminated, predominantly low density, regions remote from late
time star formation regions and a relic from
very early low mass galaxy formation.
It is stressed that the exact metallicity value of the 
the lower metallicity peak is very uncertain,
significantly subject to resolution effects;
it is likely that our still relatively poor resolution 
must have underestimated star formation at high redshift
in quite small galaxies not well resolved by our numerical resolution,
which may otherwise have enriched the IGM with metals to a higher floor level.
In addition, the metals produced by the very first generation of stars 
(e.g., Cen 2003; Ricotti \& Ostriker 2004)
may also help to have raised the metallicity floor in the IGM.
Overall, the effect of GSW is to reduce the amount of metals
in the high metallicity peak ($Z\sim 0.3$ to $1.0\zsun$) of the warm gas
and it is likely that a significant amount of this metal-rich gas is dispersed 
to mix with other lower metallicity warm gas in the 
metallicity range ($Z\sim 0.003$ to $0.3\zsun$).
More importantly here, a significant amount of warm, metal-rich
gas is heated to the WHIM temperature range,
accounting for the differences seen at 
$Z\sim 0.01$ to $1.0\zsun$ between the simulations without and with GSW
shown in the middle panel of Figure 15.
At the same time, it is also clearly seen that some of the WHIM gas
at low metallicity 
($Z\le 0.01\zsun$) is substantially enriched and hence
removed from that section and added to the section with $Z\ge 0.01\zsun$.
We find that the median metallicity of the WHIM 
is $0.18\zsun$ for oxygen with 50\% and $90\%$ intervals
being (0.040,0.38) and (0.0017,0.83).
Finally, the effect of GSW on hot, cluster X-ray gas
is rather minimal, as shown in the bottom panel of Figure 15.
The metallicity of the cluster gas is robustly peaked at $Z\sim 0.3\zsun$ 
in terms of iron abundance, as observed (Arnaud et al. 1994;
Tamura et al. 1996;
Mushotzky et al. 1996;
Mushotzky \& Lowenstein 1997; Tozzi \etal 2003);
in fact, this is how we normalize our metal yield (the sole 
parameter to determine the amount of metal output from stars)
in order to construct a self-consistent picture.
In our simplest, one-parameter model for metal yield,
there is no adjustable free parameter
that can alter the distribution of metals
and our predictions as presented in the next section are completely 
deterministic and falsifiable.

Figure 16 shows the cumulative metal mass fraction 
as a function of gas density for each of the three IGM components.
Overall, we find that in the simulation without GSW,
$(36\%,48\%,16\%)$ of all metals in IGM are in the (warm, WHIM, hot) components.
The distribution among the IGM components is altered to 
(warm, WHIM, hot)$=(12\%,71\%,17\%)$ in the simulation with GSW.
This shows the GSW reduces the metal mass in the warm IGM component
by a factor of $3$!, while the hot IGM component is virtually unaffected by GSW.
All the metals when a part of the warm IGM component
heated up are added to the WHIM component.
{\it This is the most dramatic effect that GSW have on one single
physical component/process.}
Also quite clear from the top panel of Figure 16,
most of the metals that are heated and removed from the 
warm component are in relatively overdense regions,
at $\rho_{gas}\ge 200$.
However, most of the metals that are added to the WHIM
are seen, from the middle panel of Figure 16,
to be in regions spanning a significant range in density
from $1-300$ times the mean density.
At the same time, as the middle panel of Figure 16 indicates,
some WHIM gas in high density region ($\rho > 300$) 
is affected in the opposite sense that
their metallicity appears to be reduced by GSW (that
the thick solid curve falls below the thin solid curve);
however, a more correct interpretation is that the thin solid curve
is pushed to the left to become the thick solid curve, i.e.,
GSW reduce gas density of this already metal-rich gas which is
located in the very close vicinity of galaxies.

Figure 17 further illustrates the effect of GSW on metal transport
in a ``phase" space,
where we show {\it average} gas metallicity in the density-temperature 
phase space for the simulations with (top panel) and without (bottom panel) GSW.
Let us examine them closely.
First, in the bottom panel, we see a concentration of high metallicity gas ($Z\ge \zsun$)
at $(\rho,T)=(>10^4,\sim 10^4~K)$, 
within the black contours
which clearly is the gas
w5thin the galaxy that has been enriched by metals but remains cold-warm
in the absence of GSW shocks.
We note that the original gravitational shocks caused
the gas to cool and condense to that phase space domain.
Then in the upper panel this contour has vanished,
this concentration of high-metallicity gas is essentially
``blown away" by the GSW, as there remains little 
high metallicity gas in that phase space domain in the
simulation with GSW (top panel).
but in the upper panel this gas reappears at low density
and high temperature within the black contour in the lower left corner.
Specifically, as GSW propagate to lower density regions (cf. Figures 4,5,6),
the metals transported there weigh relatively more importantly
than in high density regions, creating high metallicity gas there.
We see that in the WHIM temperature range of $10^{5-7}$~K
there is a concentration of solar metallicity gas in the density
range of $1-30$ times the mean density,
which can be easily probed by metal aborption line observations
(see Cen \& Fang 2005). 
We note that what is displayed here is the average metallicity
and there is a very large dispersion in metallicity
for gas elements in the same phase space location.
Third, a significant fraction of WHIM gas in moderate to high density
regions of $\rho=50-10^5$ is now significantly enriched to
an average metallicity of $0.1-0.3\zsun$.

From these results we expect that one of the most
sensitive tests of GSW effect may lie in the properties of
metal rich gas in the vicinity of galaxies and groups
of galaxies, where density ranges from 
moderate to high and GSW are most energetically relevant.
These dramatic effects {caused only by GSW}, {\it not gravitational shocks},
can then be used to provide important tools to understand the GSW.
Observationally, we expect metal lines, both metal absorption and emission lines,
may serve as an excellent diagnostic for probing GSW,
as will be shown in a later section.
The absorption lines may be most efficient to probe the moderate density regions,
while the emission lines may mostly concentrate on the high density regions.

\section{Discussion and Conclusions}

Numerical simulations of the intergalactic medium have shown
that at the present epoch a significant fraction ($40-50\%$) of
the baryonic component should be found in the ($T\sim 10^{5-7}$K)
Warm-Hot Intergalactic Medium (WHIM) - with 
several recent observational lines of evidence indicating the validity 
of the prediction.
We here recompute the evolution of the WHIM
with the following major improvements:
(1) galactic superwind feedback processes from galaxy/star formation 
are explicitly included;
(2) major metal species (O IV to O IX)
are computed explicitly in a non-equilibrium way;
(3) mass and spatial dynamic ranges are larger by
a factor of 8 and 2, respectively, than our previous
simulations.

Our significantly improved simulations confirm 
previous conclusions based on earlier simulations:
{\it nearly one half of all baryons at the present epoch
should be found in the WHIM} - a filamentary network 
in the temperature range of $10^5-10^7$~K.

Here are the major findings:
(1) Overall, the fraction of baryons in WHIM 
is slightly increased from the earlier work but consistent
with the fraction at $\sim 40-50\%$.
(2) galactic superwinds have three significant effects, 
first by increasing the WHIM mass fraction by about $20\%$, 
through shock heating photoionized gas
adjacent to filaments, second by 
contaminating nearby gas with extra metals and additional heating,
and third by dispersing much of the metals within galaxies.
(3) the gas density of the WHIM is broadly peaked at a density $10-20$ times the mean density
ranging from underdense regions to regions overdense by $10^3-10^4$.
(4) the median metallicity of the WHIM 
is $0.18\zsun$ for oxygen with 50\% and $90\%$ intervals
being $(0.040,0.38)\zsun$ and $(0.0017,0.83)\zsun$.

The physics behind this robust conclusion 
is largely dictated by the dominance of gravitational
heating when large-scale structures begin to 
collapse in the recent past in the cosmic history.
Simply put, the average temperature of 
the intergalactic medium at any epoch is determined
by the mass scale that goes nonlinear at that epoch,
which, in present time, is close to the $8h^{-1}$Mpc
scale which fixes the abundance of rich clusters.
This explains the relative insensitivity of the computed
mass fraction in the WHIM on cosmological parameters,
so long as each model is properly normalized to
reproduce the well determined abundance of observed rich clusters
of galaxies today;
only a very small extrapolation
is needed to go from this well observed scale to the nonlinear scale,
so the estimated temperature of collapsed regions
that we find will be common to all models based on the
gravitational growth of structure - 
as normalized to local cluster observations.

It was not entirely clear how important feedback processes
from galaxy formation on IGM are, based on our early simulations
compared to the simulations by others (Dav\'e \etal 2001).
In this work we demonstrate quantitatively, for the first time,
this effect.
We find that galactic superwinds generated collectively
by supernova explosions in galaxies, while still subdominant to
gravitational heating of large-scale structure formation,
have important effects on the WHIM as well as other components
of the IGM.
The mass fraction of the WHIM is increased by about 20\%,
when GSW are included.
We show that this increase in the WHIM mass fraction comes
largely at the expense of the warm IGM phase over 
a broad range $30-10^4$ times the mean density,
which is often independently invoked to suppress
star formation to alleviate the overcooling problem.
A perhaps more important effect of GSW is to transport
metals to the IGM from inside galaxies to a distance
of several hundred kiloparsecs up to about one megaparsec,
a necessary process to enrich the IGM. 
It is therefore likely that detailed, joint analyses and
comparisons with observations of the detailed structure
of the WHIM density, temperature and metal abundances
should be able to provide 
useful information on the GSW, which currently can only
be modeled crudely.
Nevertheless, with the adopted approximate treatment of GSW,
we are able to
show the significant effects on many observables, 
including the properties of 
major absorption lines and emissions lines in UV and soft X-ray.

\acknowledgments

We thank Taotao Fang, Ed Jenkins and Mike Shull for useful comments.
The simulations were performed at
the Pittsburgh Supercomputer Center.
We thank R. Reddy at 
the Pittsburgh Supercomputer Center
for constant and helpful assistance 
in the process of making the simulations.
This work is supported in part by grants NNG05GK10G
and AST-0507521.

\begin{figure}
\plotone{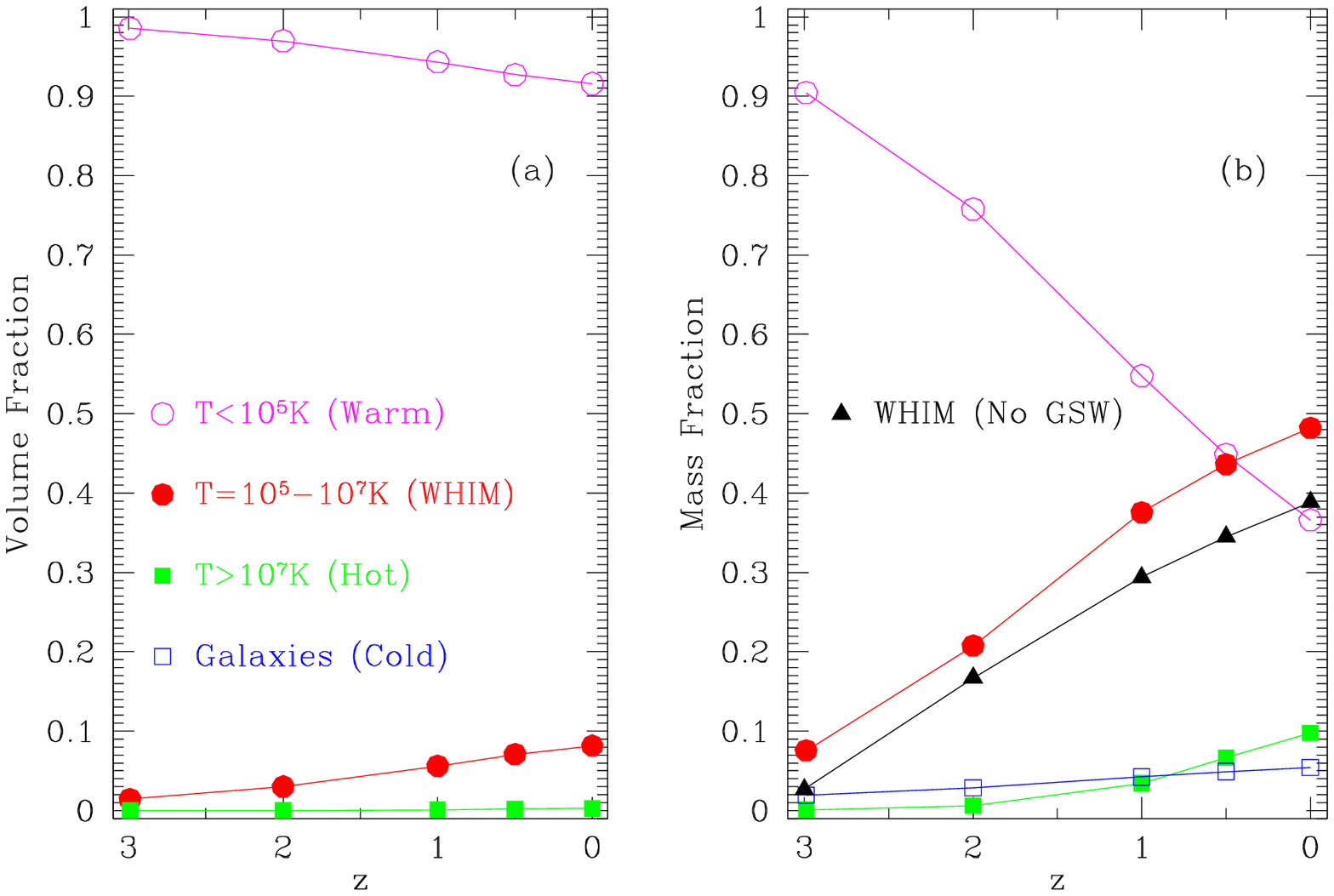}
\vskip -4cm
\caption{
shows the evolution of the four components of cosmic baryons
(see text for definitions) for the simulation
with GSW.
Panel (a) shows the volume fractions of the four components
and 
panel (b) shows the mass fractions.
Examination of (b) shows that about one half
of all baryons at redshift zero are in the WHIM.
Also included in Panel (b) as solid triangles
is the mass evolution of the WHIM from the simulation with no GSW.
}
\label{f1}
\end{figure}

\clearpage
\begin{figure}
\plotone{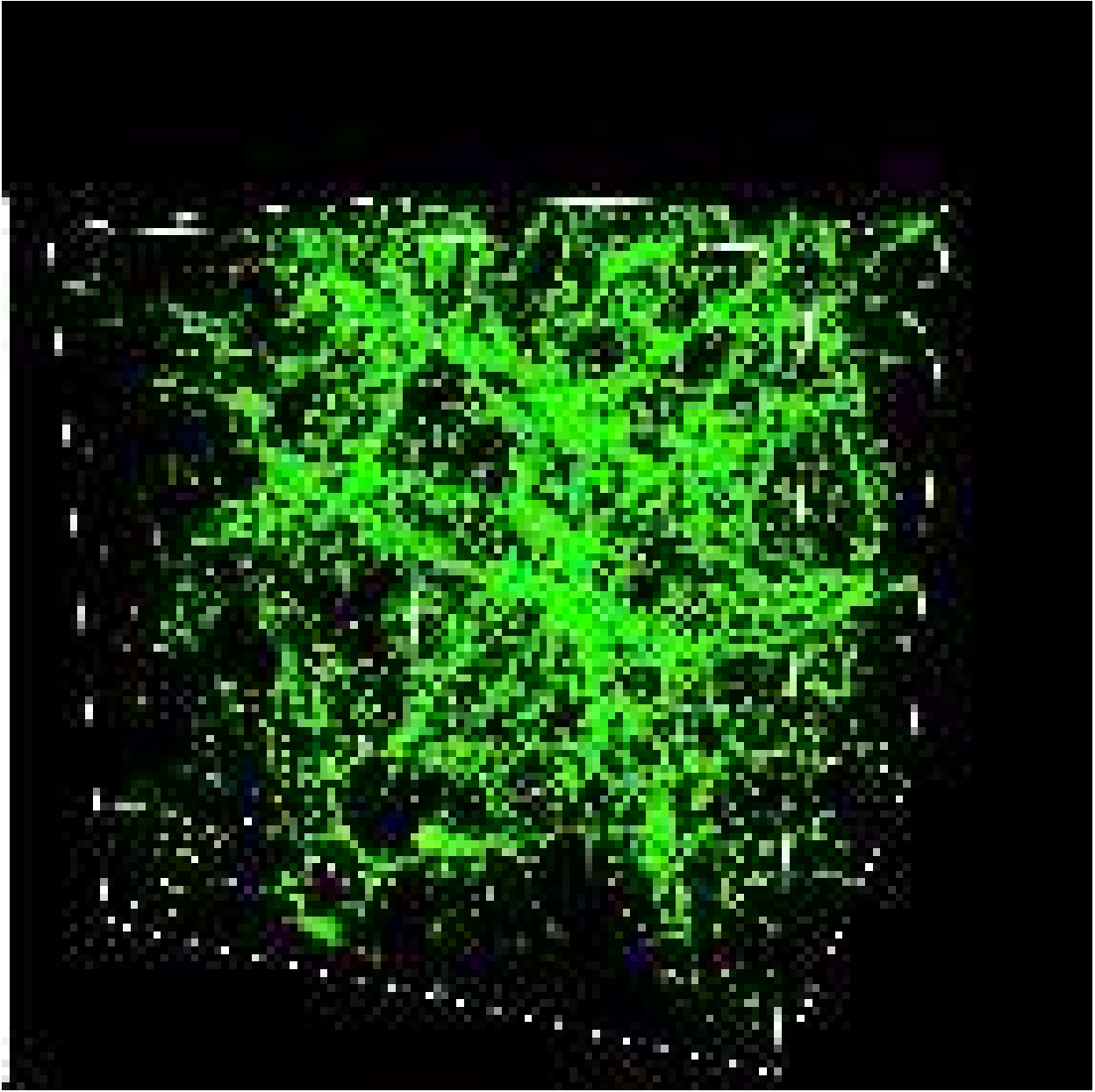}
\caption{
shows the spatial distribution of the
warm/hot gas with temperature in the range
$10^{5-7}$K at $z=0$ in a box of size $85h^{-1}$Mpc/h containing
$10^9$ individual cells,
showing the characteristic appearance of a ``Cosmic Web".
The green regions have densities about
$10-20$ times the mean baryon density of the universe at $z=0$;
the yellow regions have densities about one hundred
times the mean baryon density, while
the small isolated regions with red and saturated dark colors
have even higher densities reaching about
a thousand times the mean baryon density,
and are sites for current galaxy formation.
}
\label{f2}
\end{figure}

\clearpage
\begin{figure}
\includegraphics[angle=0.0,scale=0.60]{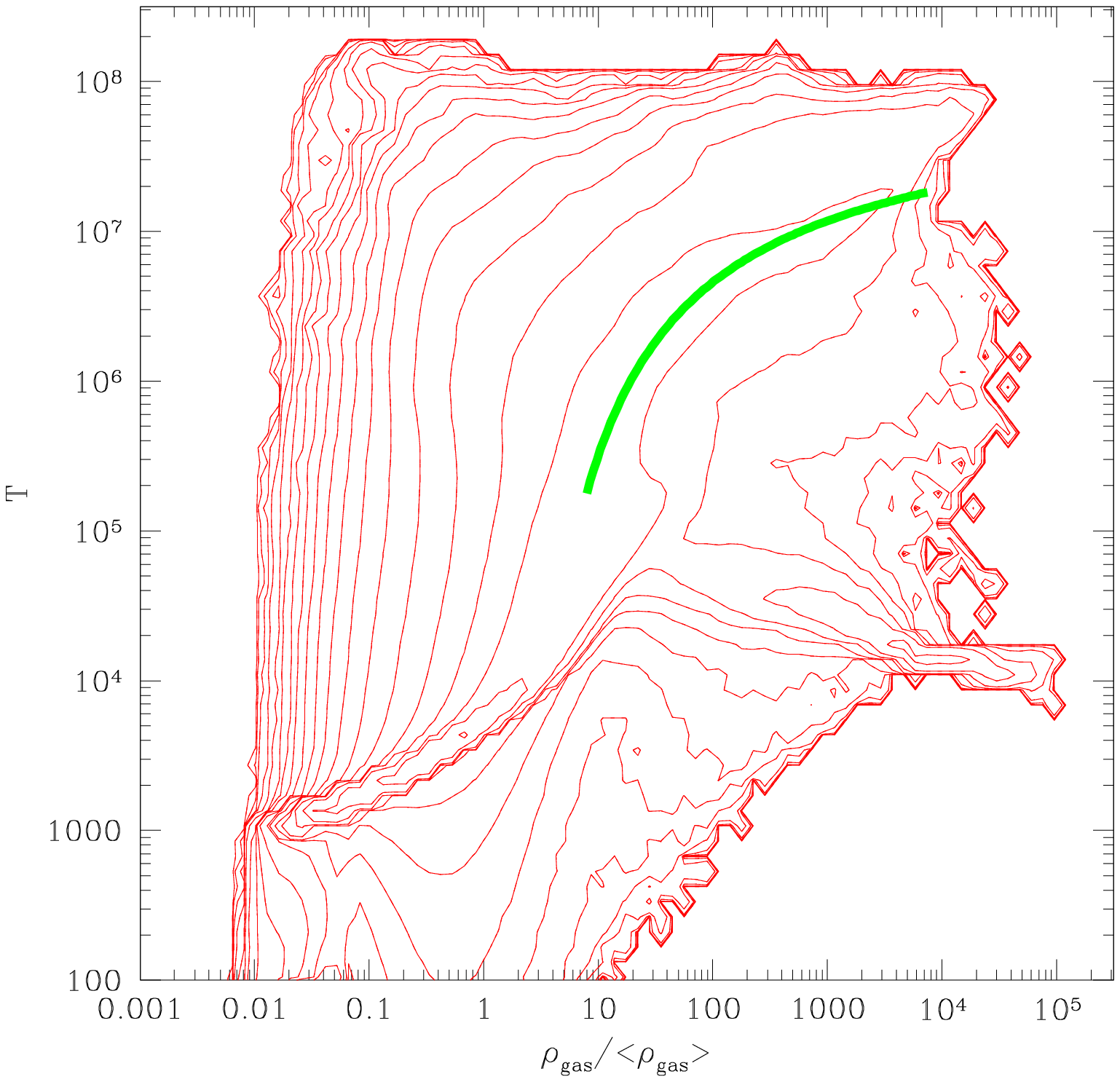}
\vskip -2.5cm
\includegraphics[angle=0.0,scale=0.60]{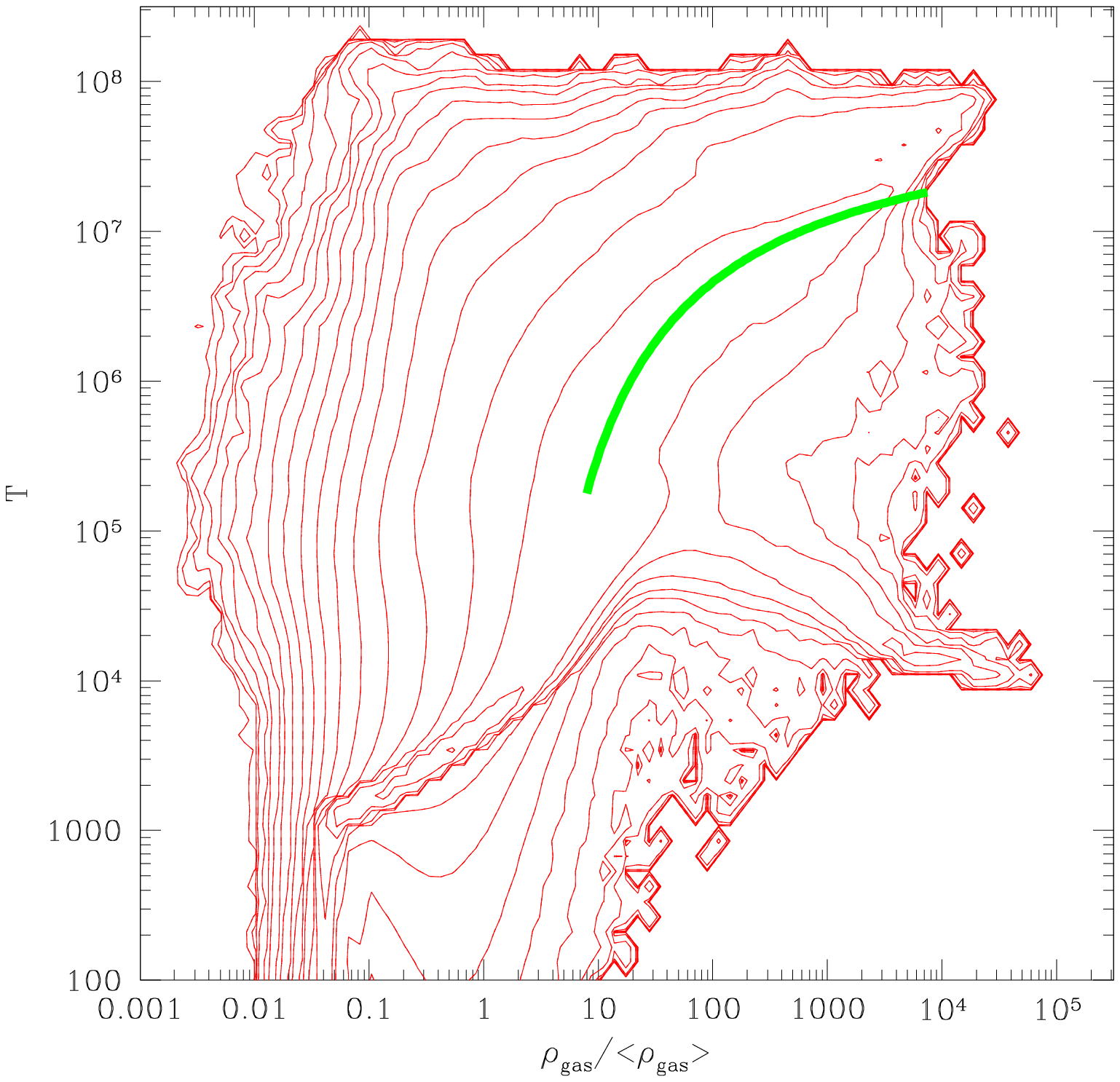}
\vskip -1.5cm
\caption{
shows the concentration of (all) gas mass in the density-temperature plane
for the run without galactic superwinds (top panel)
and with (bottom panel).
The thick solid curve is a fitting formula
tracing the locus for the WHIM ridge line (see Equation 1 in text).
There are two contour levels per decade.
}
\label{f3}
\end{figure}

\clearpage
\begin{figure}
\vskip -2.5cm
\includegraphics[angle=0.0,scale=0.55]{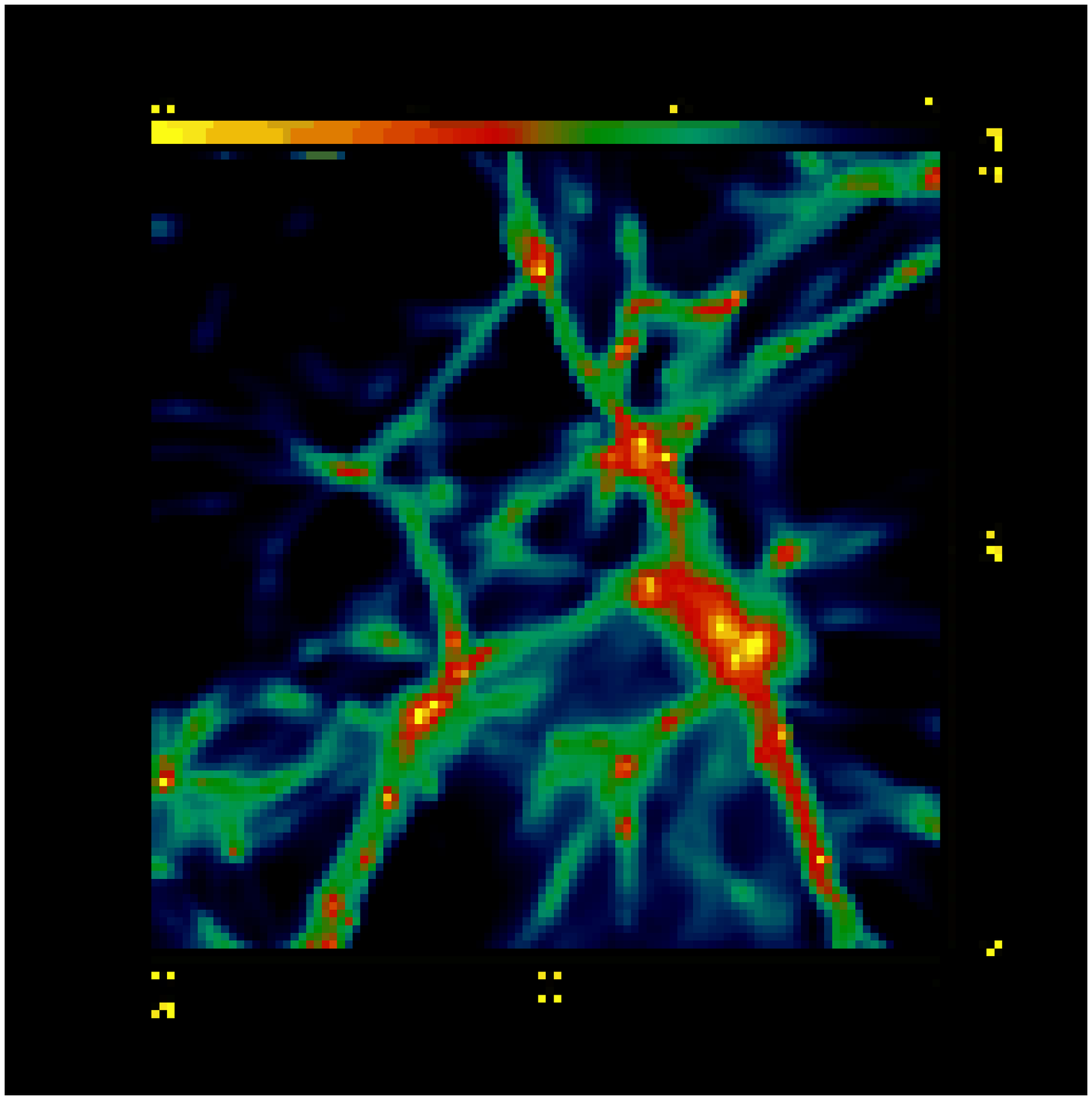}
\vskip  0.0cm
\includegraphics[angle=0.0,scale=0.55]{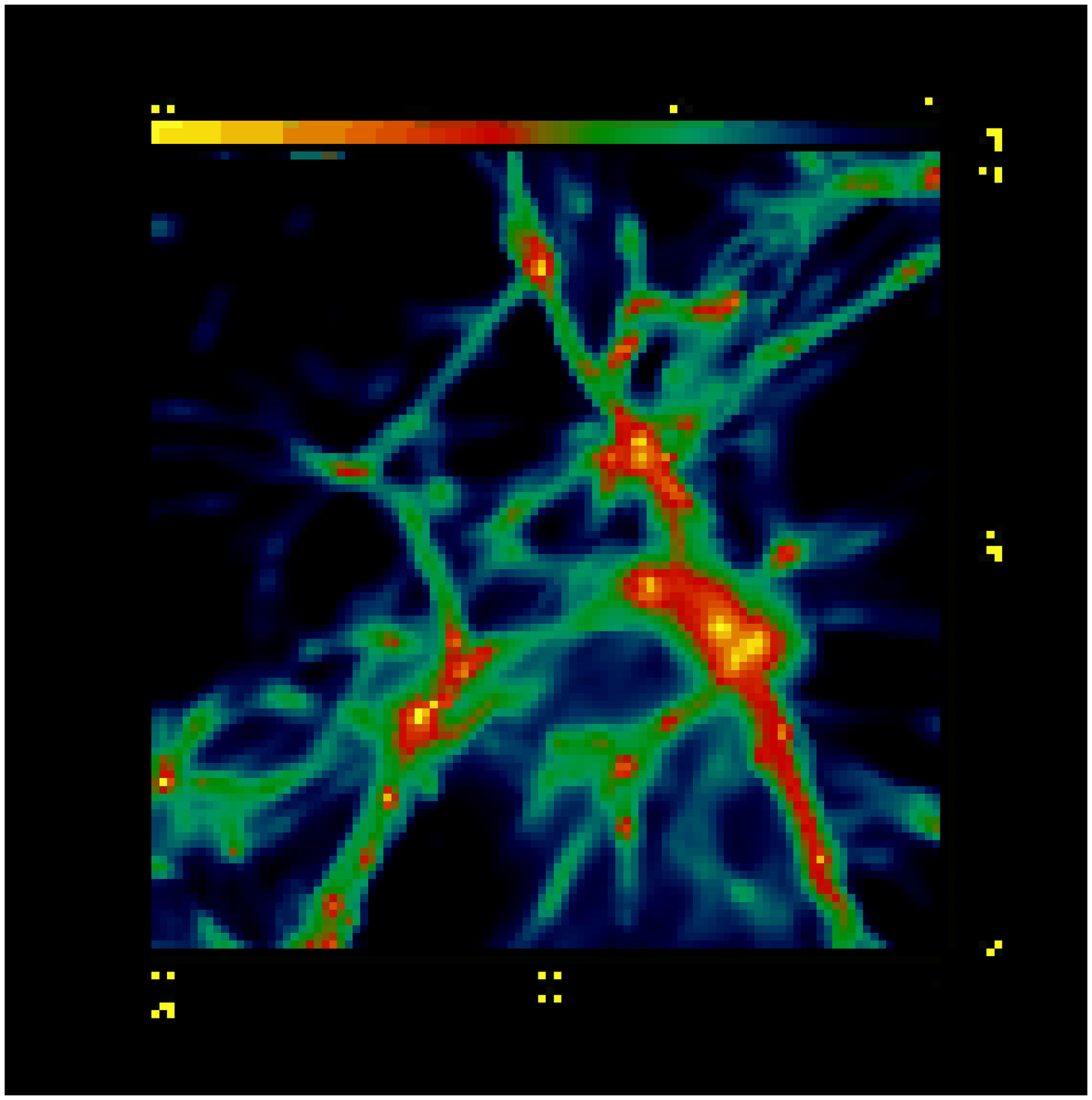}
\caption{
shows the spatial distribution of $\log(\rho){gas}$ in a thin slice
of size $21.2\times 21.2$Mpc$^2$/h$^2$
and depth $1.75$Mpc/h without (top) and with (bottom) GSW. 
Note that gas of all temperatures is included in this plot.
Note reduction of yellow high density regions as GSW push gas out to
lower  densities.
Visible examples
include the features at $(8,5)$Mpc/h, $(13,8)$Mpc/h and $(6,13)$Mpc/h.
}
\label{f2}
\end{figure}

\clearpage
\begin{figure}
\vskip -2.5cm
\includegraphics[angle=0.0,scale=0.55]{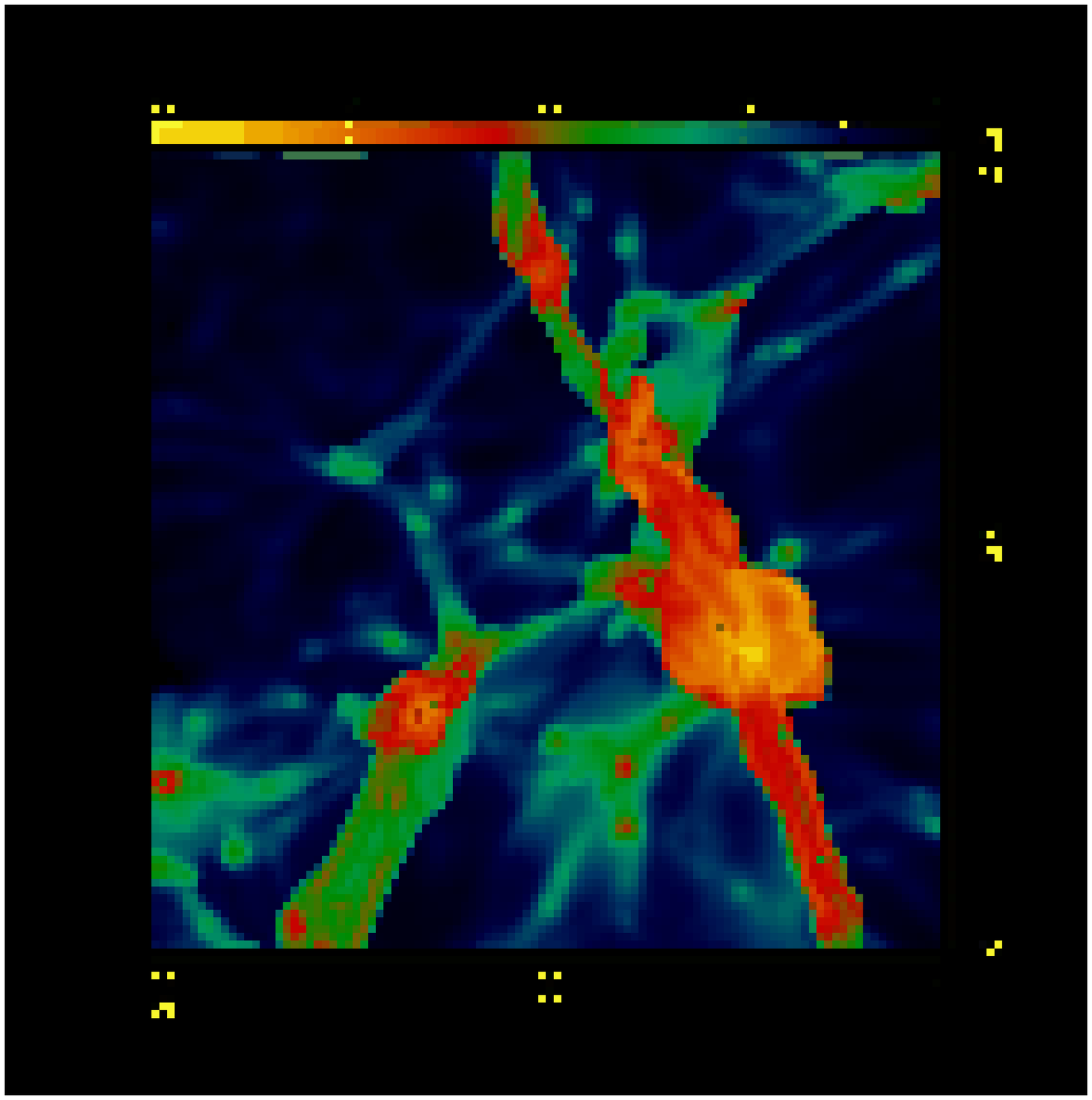}
\vskip  0.0cm
\includegraphics[angle=0.0,scale=0.55]{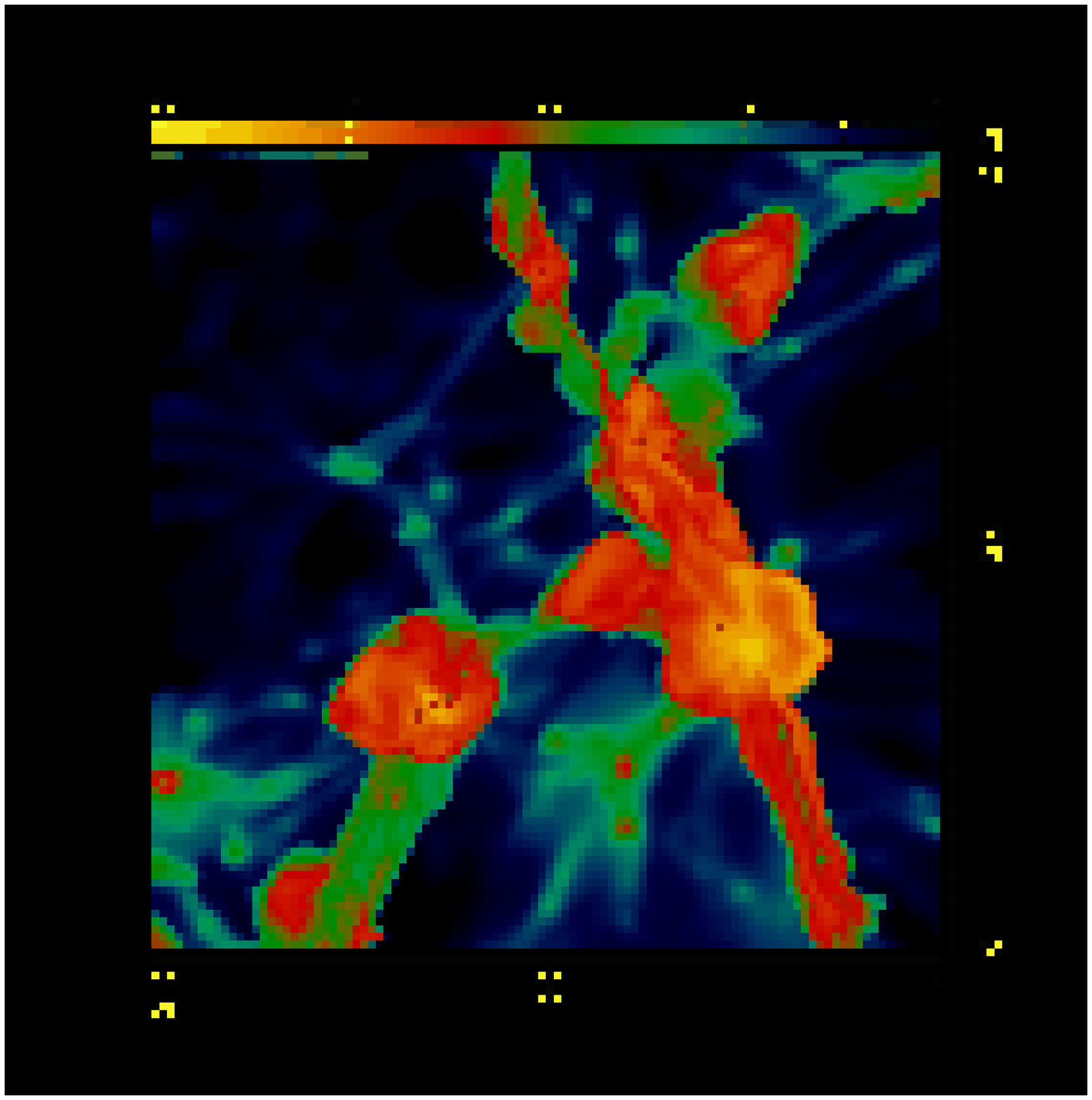}
\caption{
shows the spatial distribution of $\log$(gas temperature) in the same slice
without (top) and with (bottom) GSW. 
The temperature is units of Kelvin.
Note shocked spurs propagating perpendicular to filaments.
}
\label{f2}
\end{figure}

\clearpage
\begin{figure}
\vskip -2.5cm
\includegraphics[angle=0.0,scale=0.55]{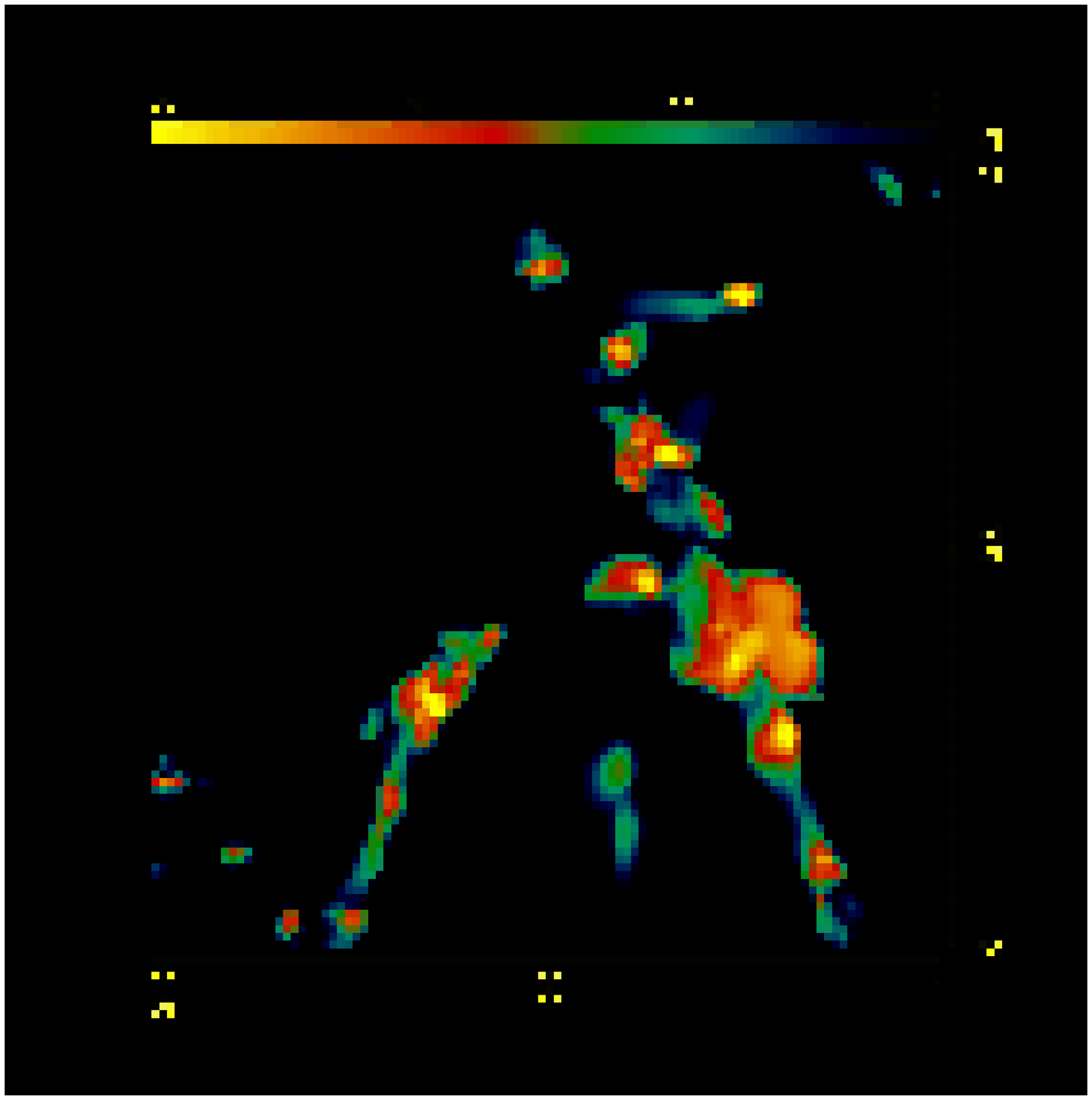}
\vskip  0.0cm
\includegraphics[angle=0.0,scale=0.55]{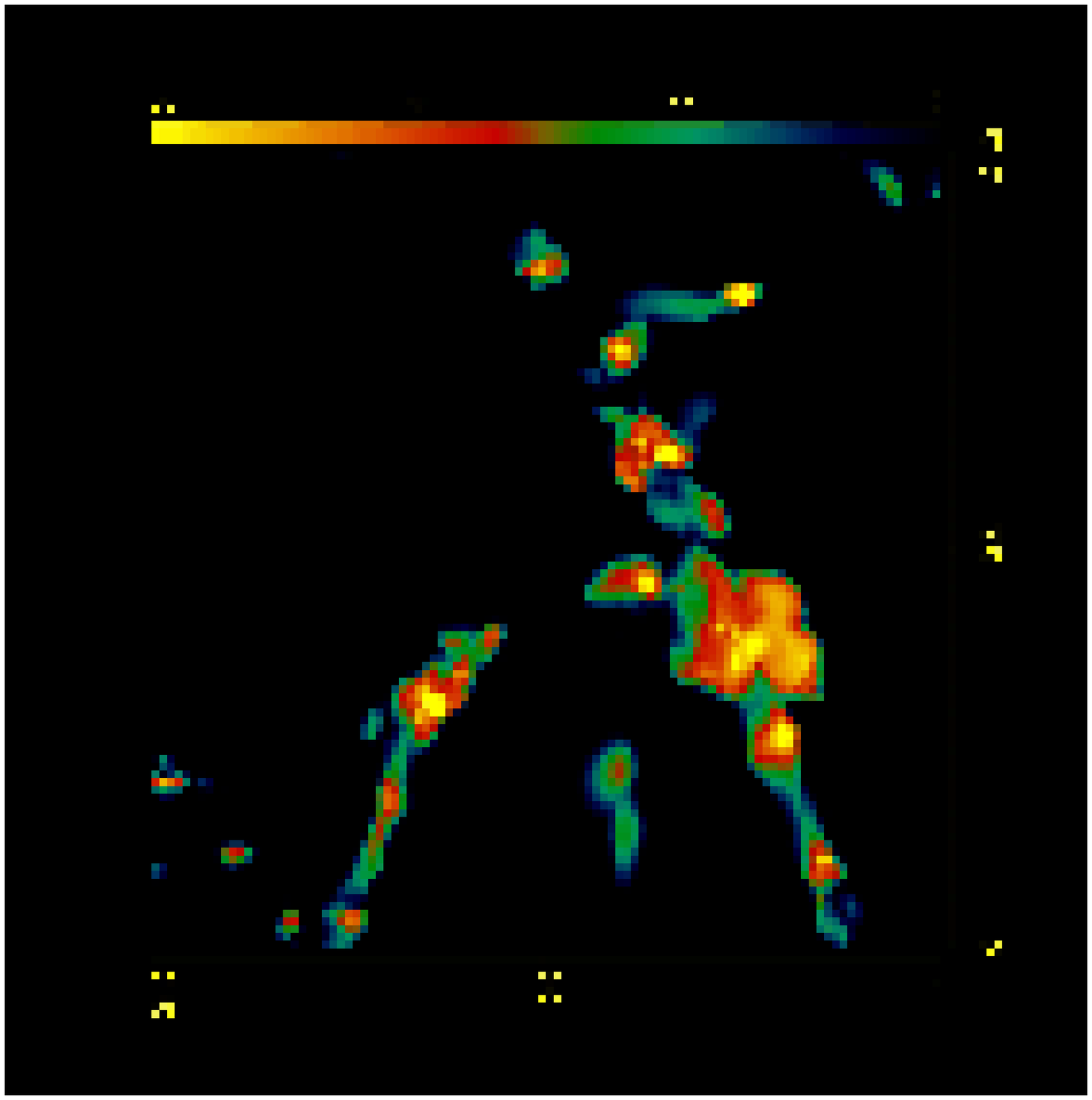}
\caption{
shows the spatial distribution of $\log$(metal density) in the same slice
without (top) and with (bottom) GSW. 
The gas metallicity is in units of solar metallicity.
Note slightly greater extant of metal rich regions with GSW.
}
\label{f2}
\end{figure}

\clearpage
\begin{figure}
\vskip -2.5cm
\includegraphics[angle=0.0,scale=0.55]{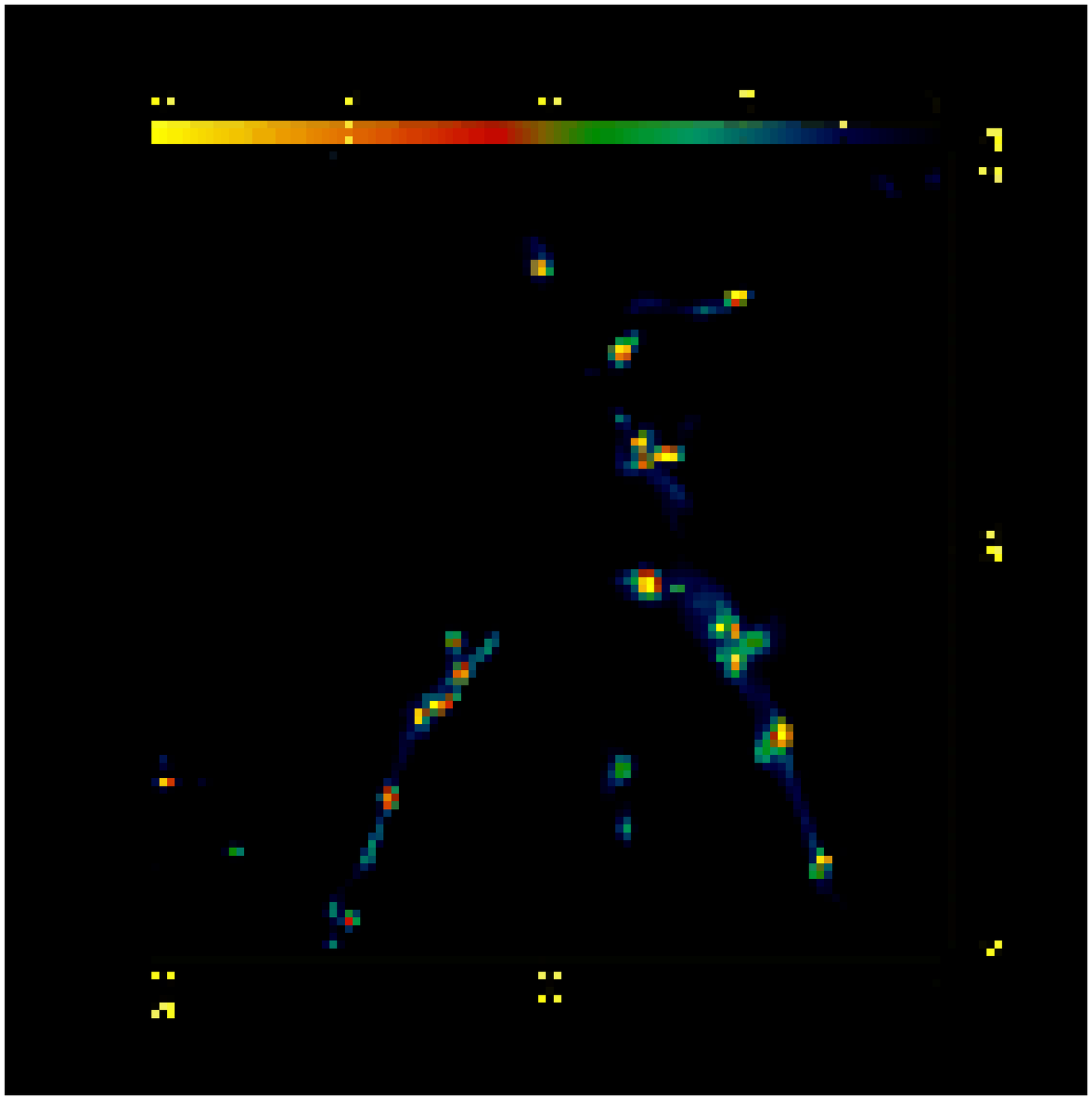}
\vskip  0.0cm
\includegraphics[angle=0.0,scale=0.55]{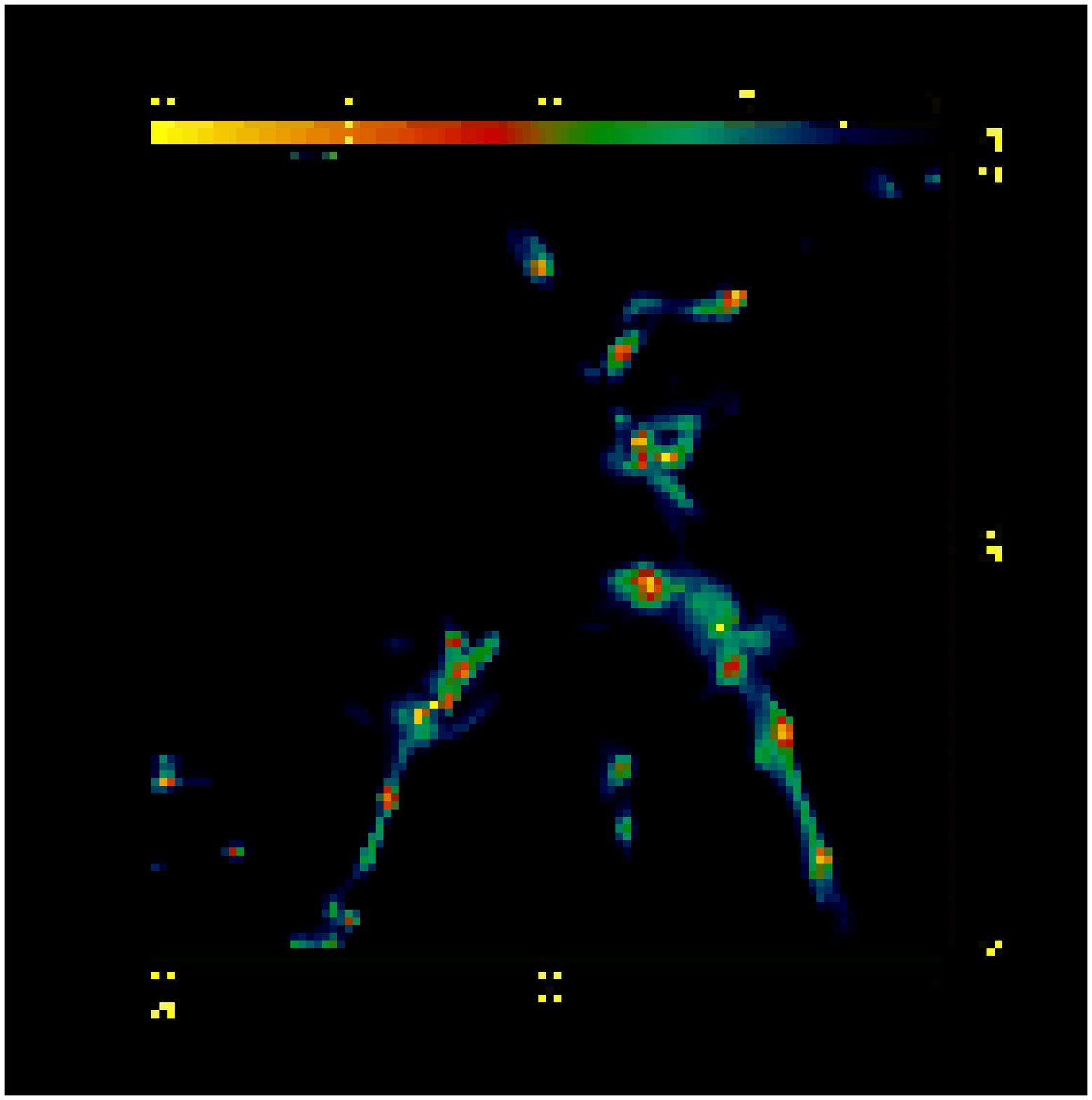}
\caption{
shows the spatial distribution of $\log$(O VI density) in the same slice
without (top) and with (bottom) GSW. 
The density is in units of the global mean gas density.
Significantly greater volume of  O VI 
reflects both temperature and metallicity effects of GSW.
}
\label{f2}
\end{figure}

\clearpage
\begin{figure}
\vskip -2.5cm
\includegraphics[angle=0.0,scale=0.55]{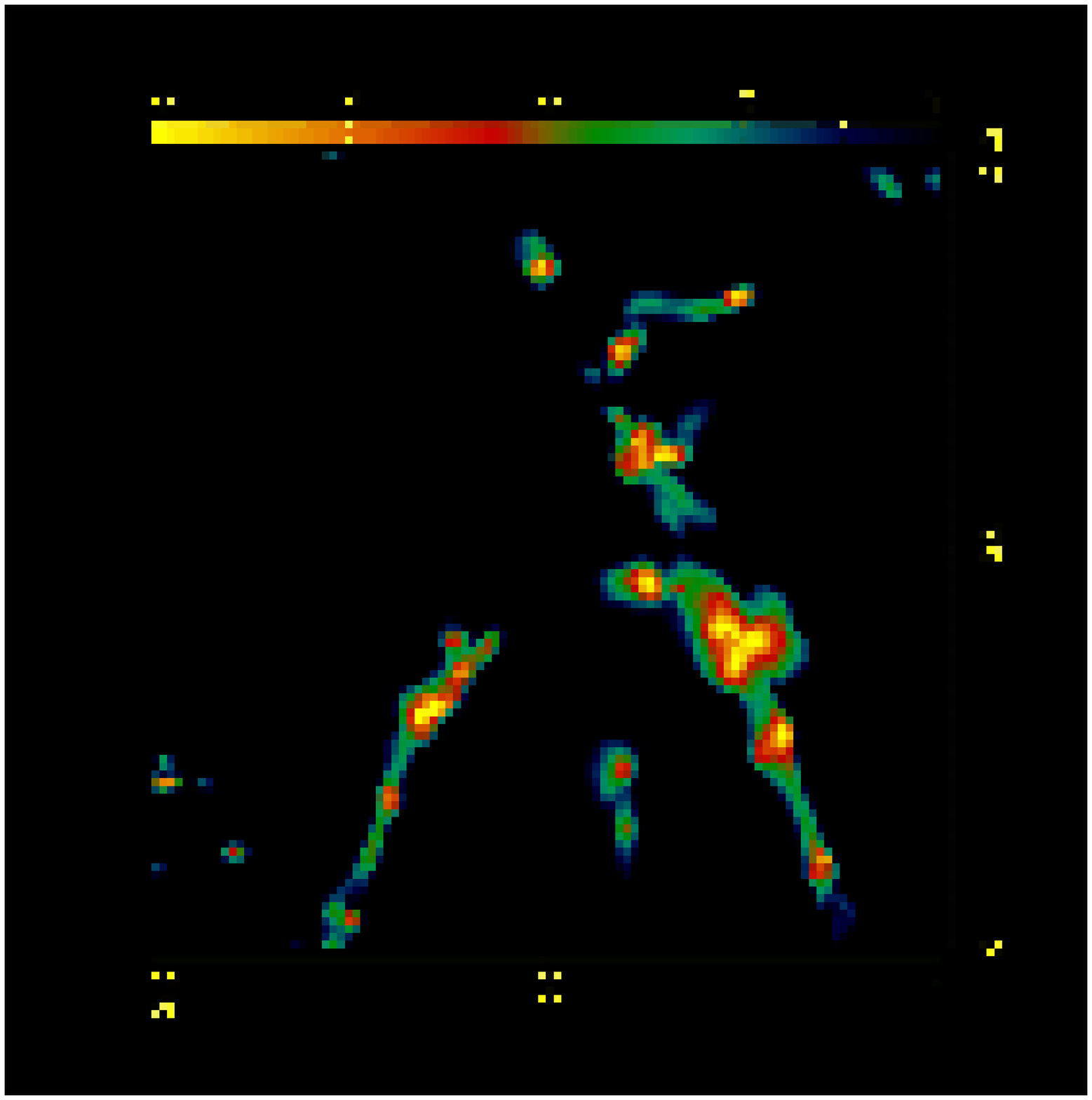}
\vskip  0.0cm
\includegraphics[angle=0.0,scale=0.55]{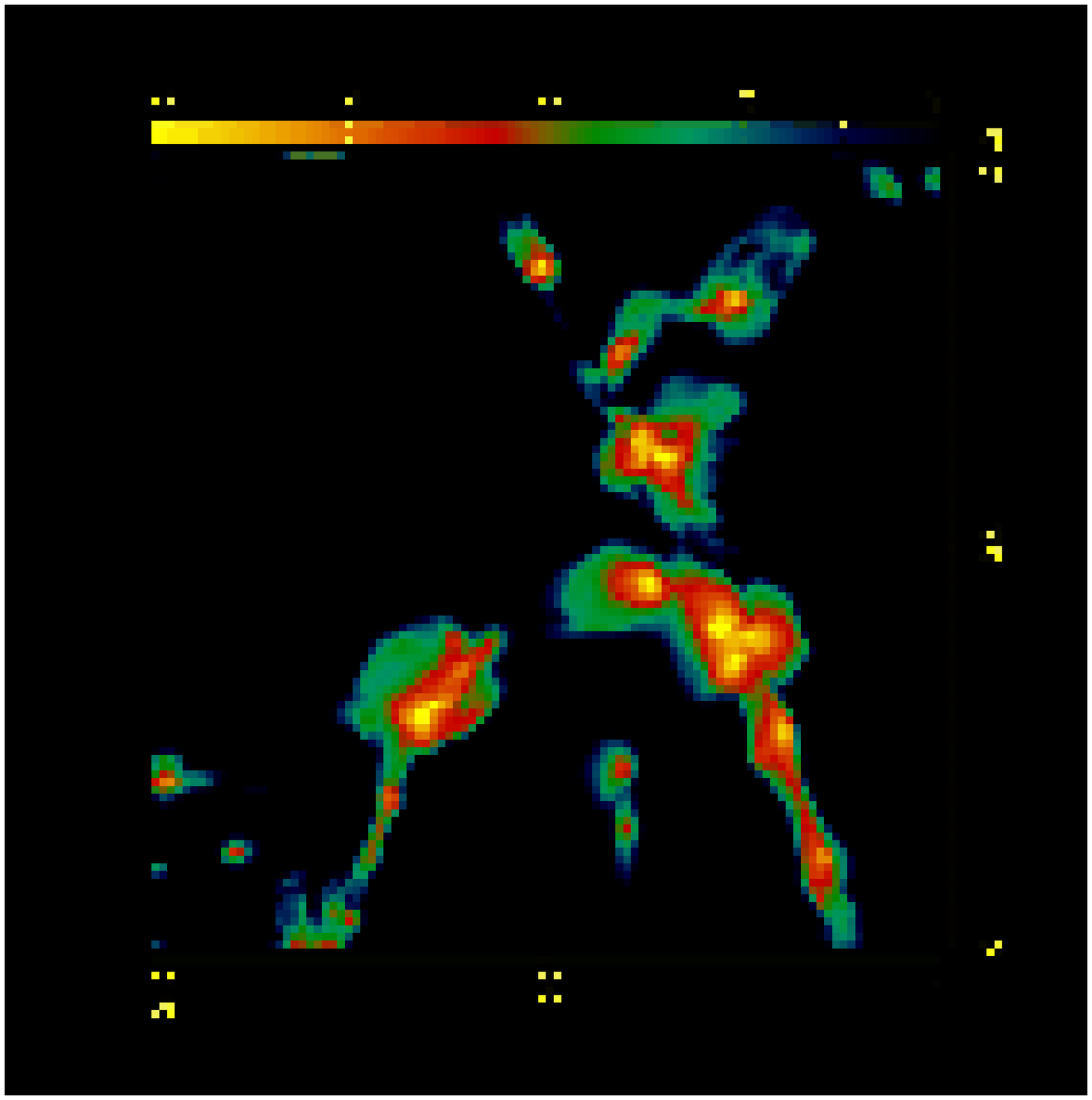}
\caption{
shows the spatial distribution of $\log$(O VII density) in the same slice
without (top) and with (bottom) GSW. 
The density is in units of the global mean gas density.
GSW effects are stronger for O VII than O VI.
}
\label{f2}
\end{figure}

\clearpage
\begin{figure}
\vskip -2.5cm
\includegraphics[angle=0.0,scale=0.55]{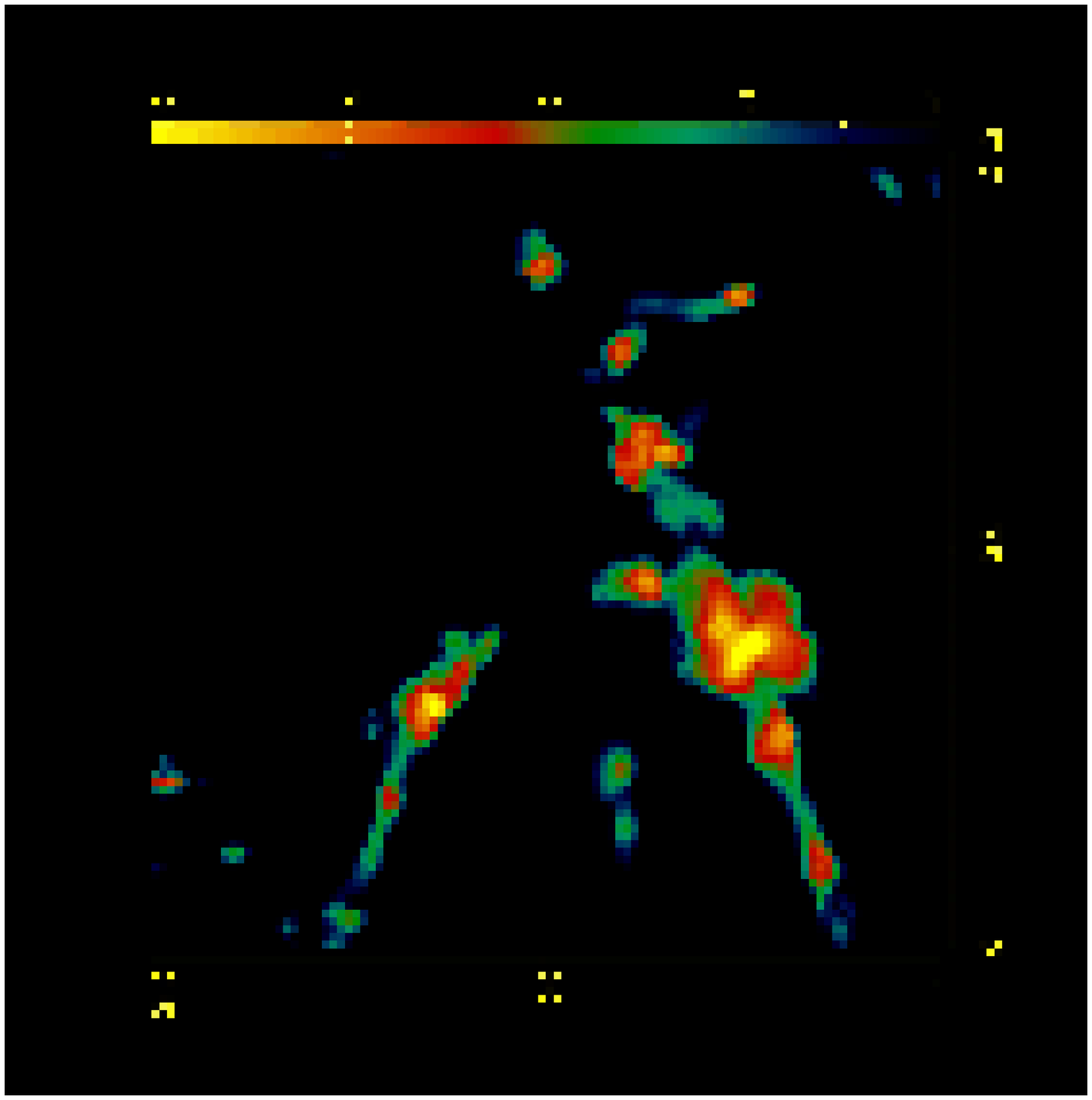}
\vskip  0.0cm
\includegraphics[angle=0.0,scale=0.55]{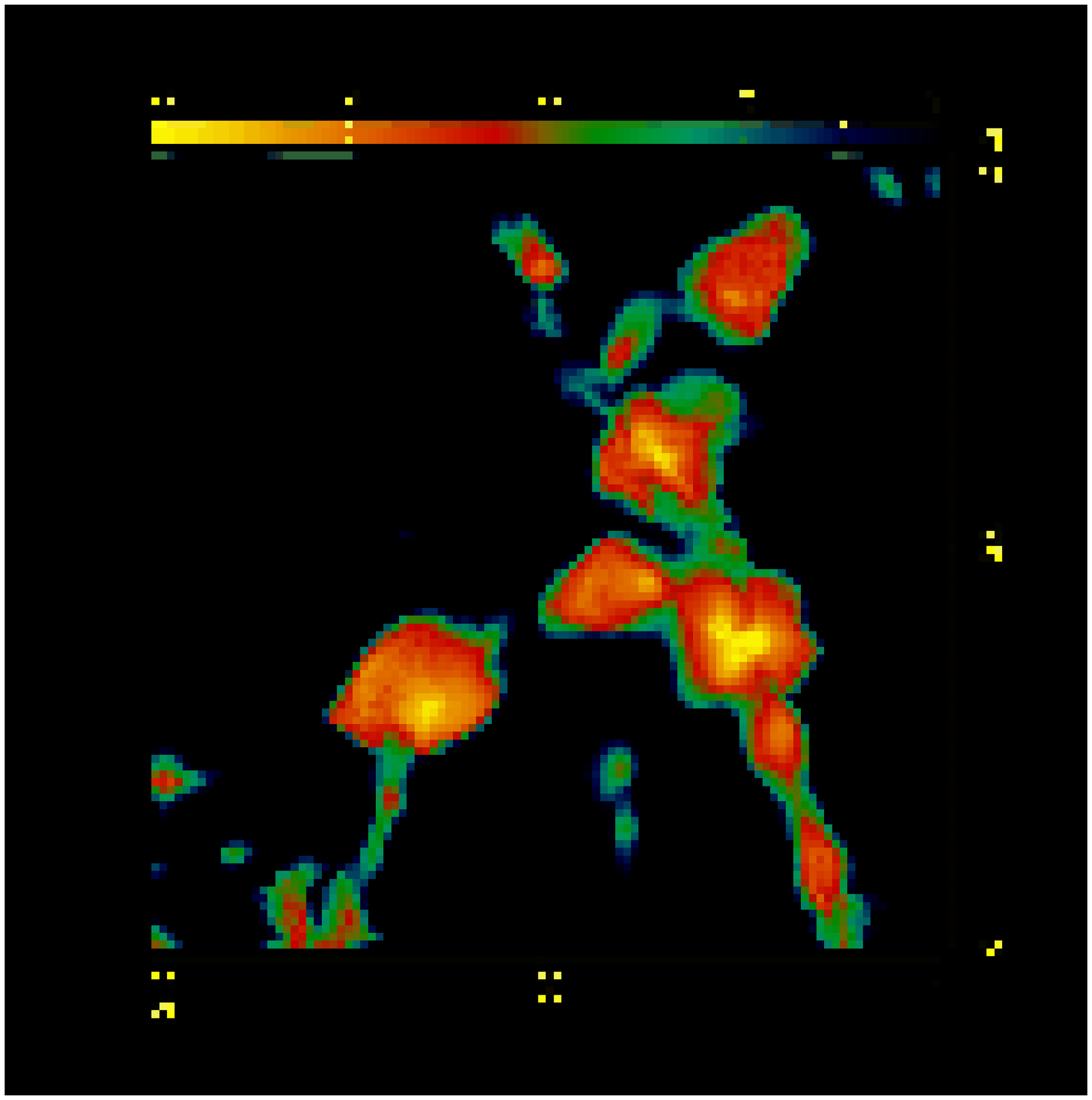}
\caption{
shows the spatial distribution of $\log$(O VIII density) in the same slice
without (top) and with (bottom) GSW. 
The density is in units of the global mean gas density.
Here GSW effects on the prevalence of O VIII are seem to be dramatic.
}
\label{f2}
\end{figure}

\begin{figure}
\plotone{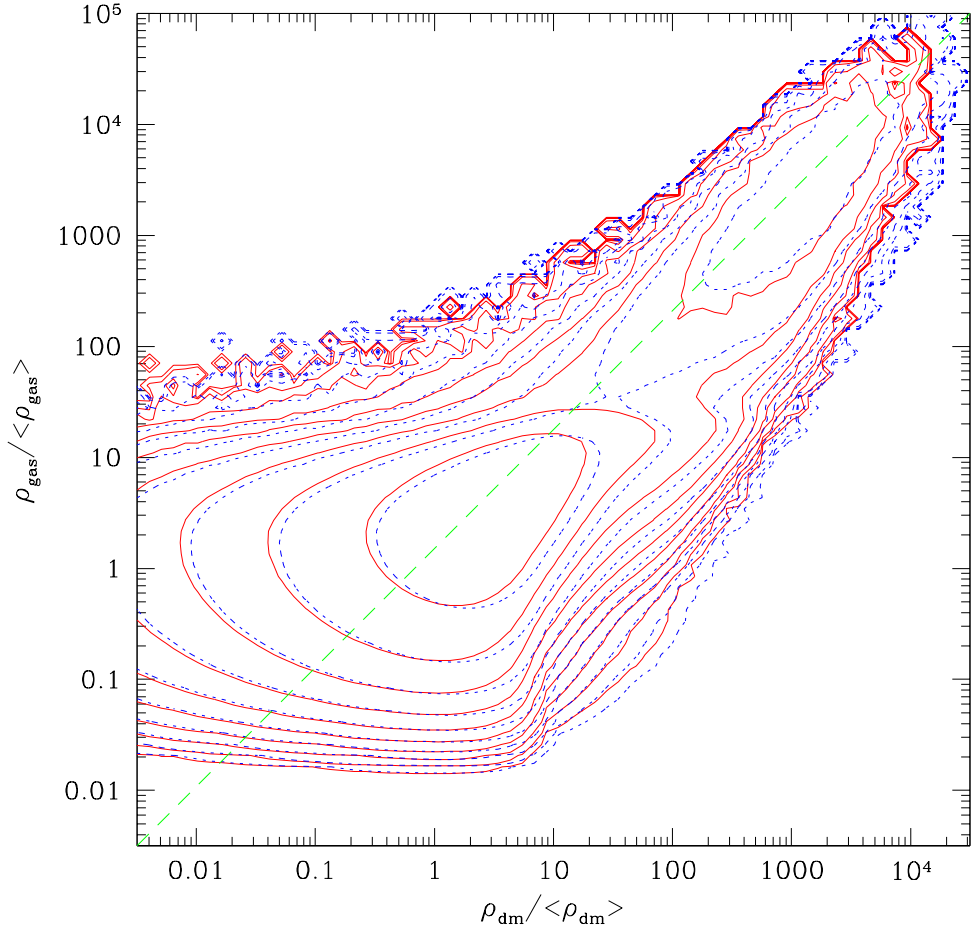}
\caption{
shows the mass concentration 
of cold gas ($T<10^5$~K) 
in the 
dark matter density-gas density plane
for the simulation with
galactic superwinds (red solid contours)
and that without galactic superwinds (blue dotted contours).
Both dark matter density and gas density are in units of 
their respective mean density.
There are two contour levels per decade.
}
\label{f5}
\end{figure}

\begin{figure}
\plotone{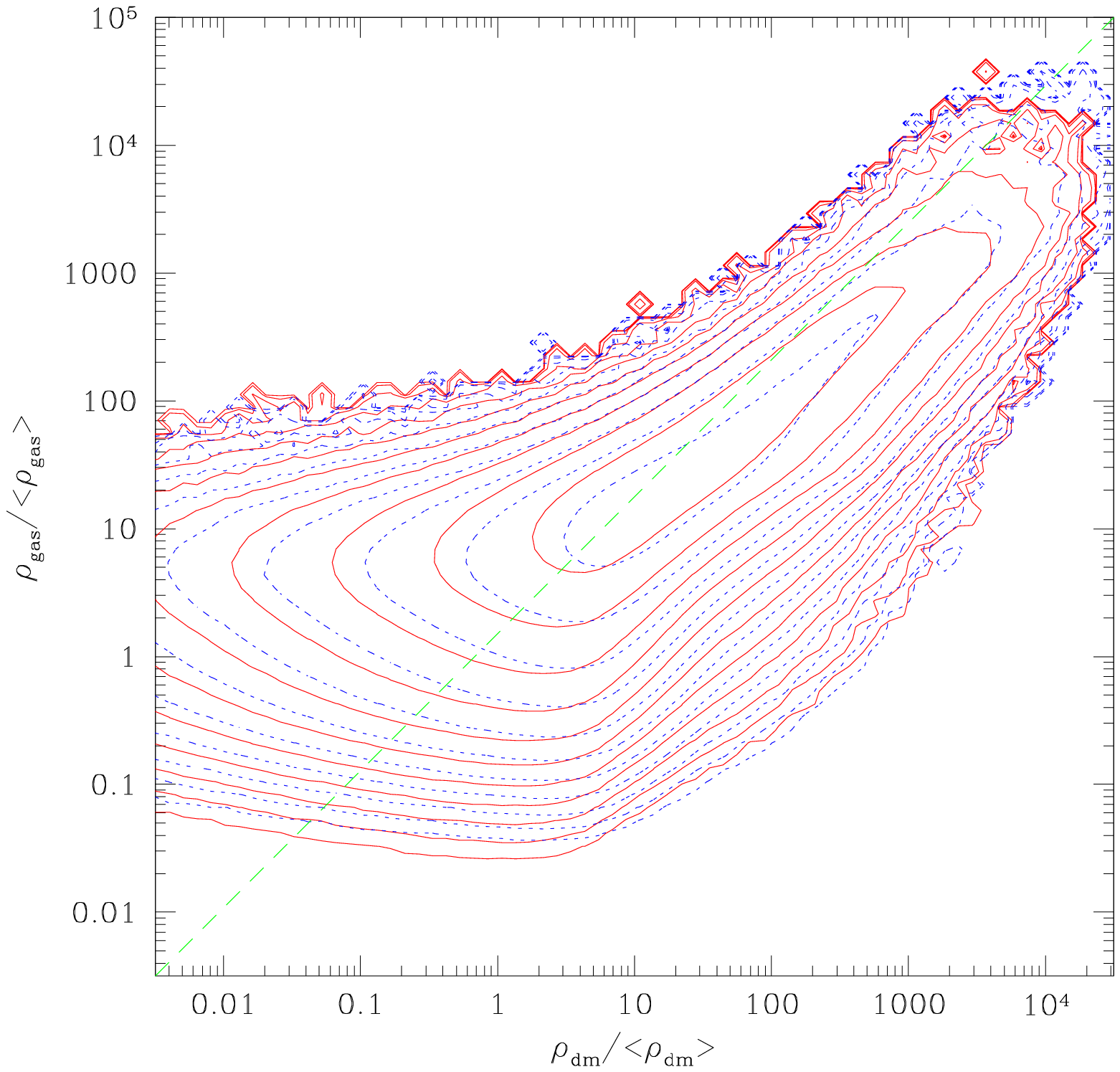}
\caption{
shows the mass concentration 
of the WHIM gas ($T=10^{5-7}$~K) 
in the dark matter density-gas density plane
for the simulation with
galactic superwinds (red solid contours)
and that without galactic superwinds (blue dotted contours).
Both dark matter density and gas density are in units of 
their respective mean density.
}
\label{f5}
\end{figure}

\begin{figure}
\plotone{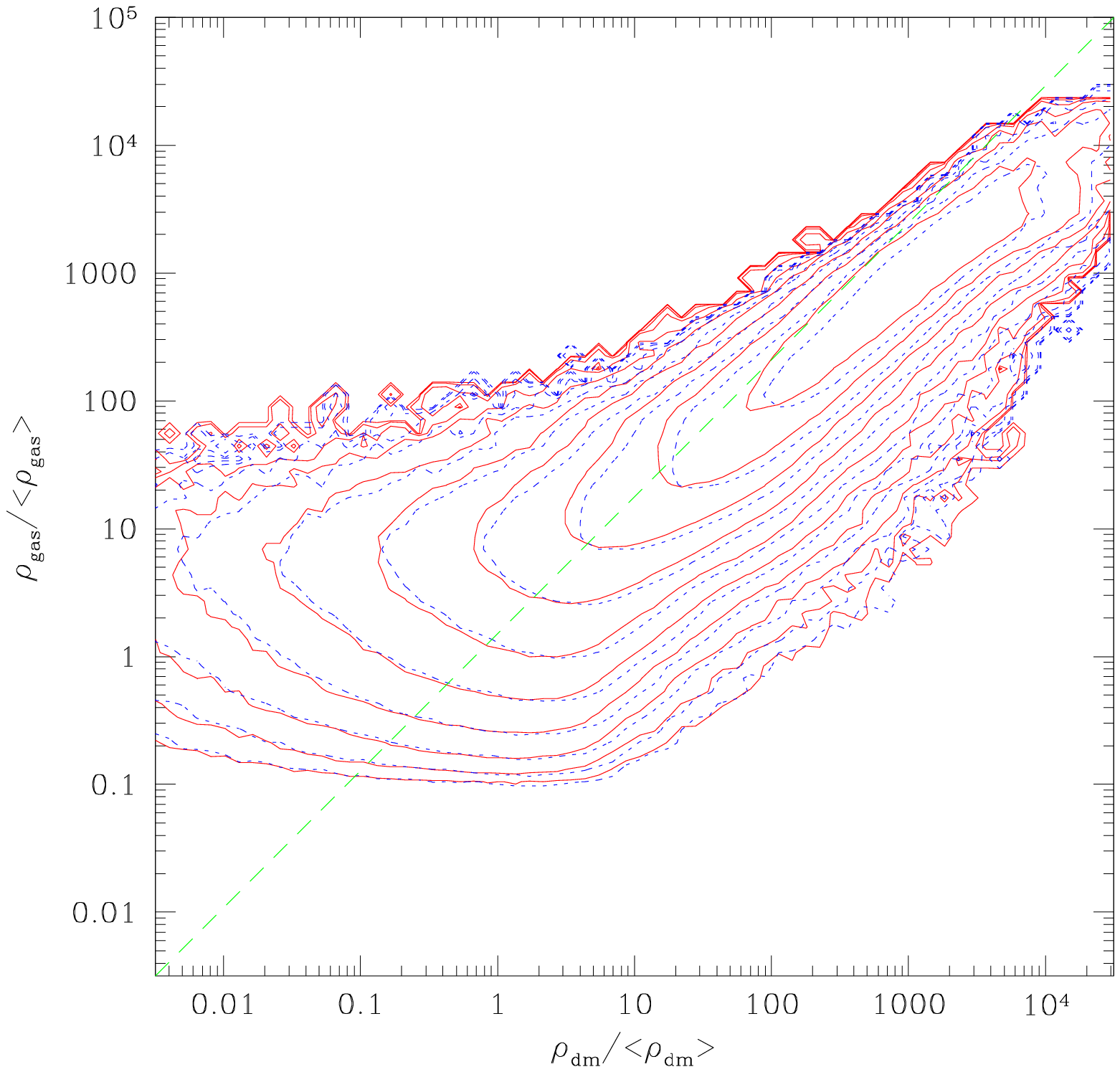}
\caption{
shows the mass concentration 
of the hot X-ray gas ($T>10^7$~K) 
in the dark matter density-gas density plane
for the simulation with
galactic superwinds (red solid contours)
and that without galactic superwinds (blue dotted contours).
Both dark matter density and gas density are in units of 
their respective mean density.
}
\label{f5}
\end{figure}

\begin{figure}
\plotone{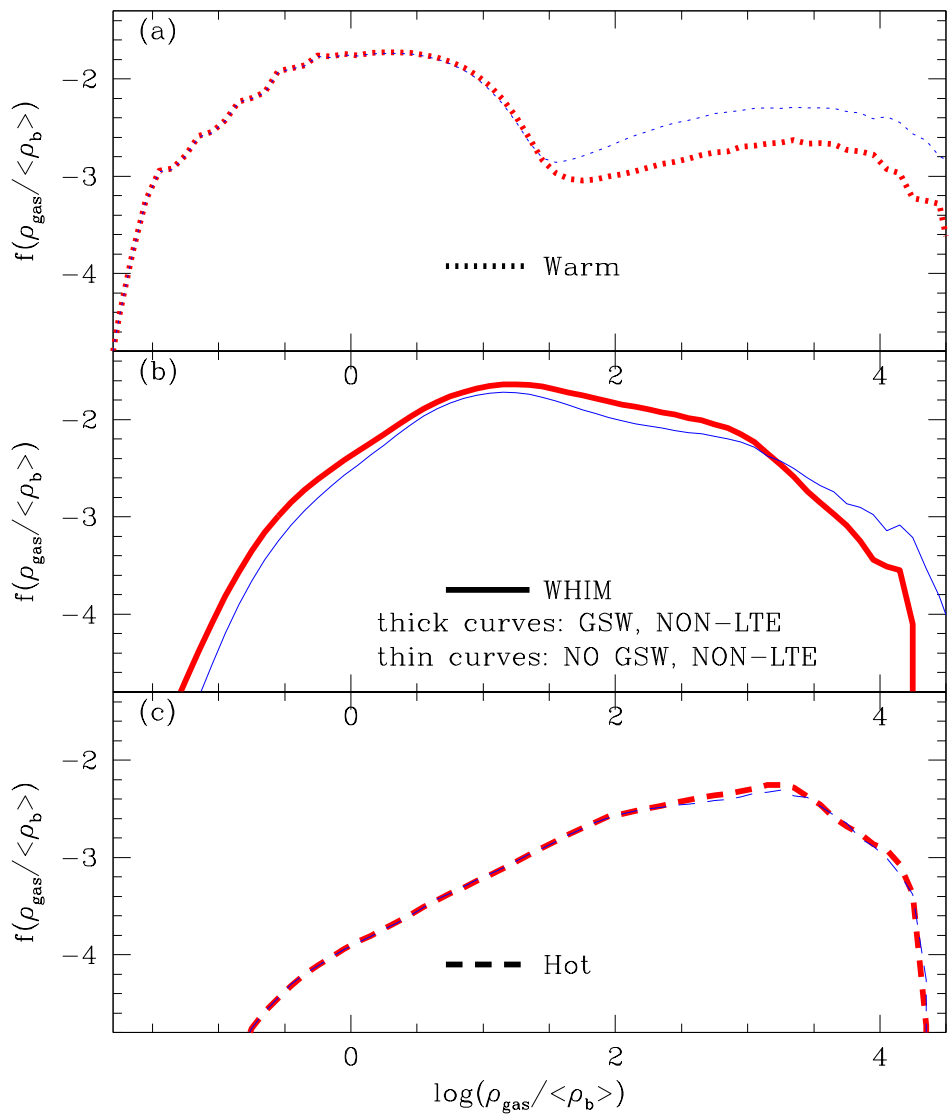}
\caption{
shows the differential 
mass fraction of three IGM components -
warm gas (dotted), WHIM (solid) and hot gas (dashed)
as a function of gas temperature 
for the simulation with
galactic superwinds (thick curves)
and that without galactic superwinds (thin curves).
The primary effect of GSW is deplete high density warm gas near
galaxies, pushing it to lower densities and higher temperatures.
}
\label{f5}
\end{figure}

\begin{figure}
\plotone{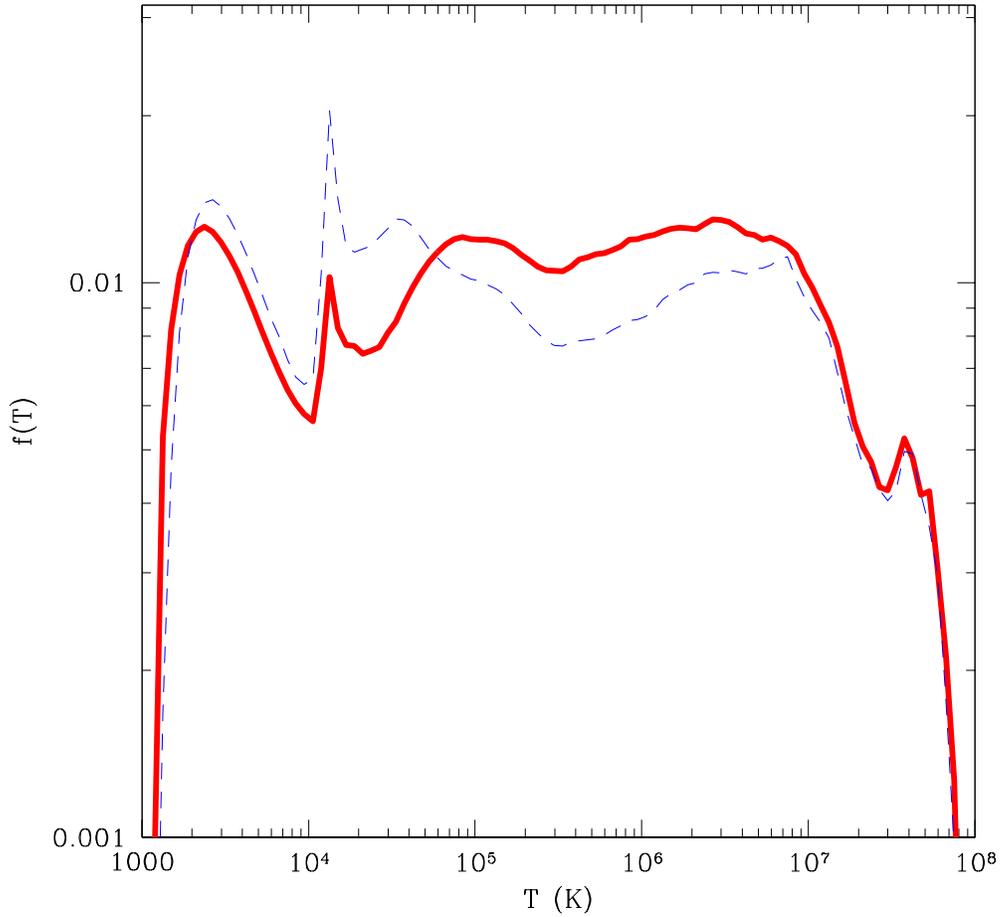}
\caption{
shows the differential 
mass fraction vs gas temperature 
for the simulation with
galactic superwinds (thick solid curves)
and that without galactic superwinds (thin dashed curves).
We see how gas in the photoionized or warm peak
at $T\sim 1.5\times 10^4$~K has been depleted and 
shock heated mostly into the $10^{5-7}$~K range.
}
\label{f5}
\end{figure}

\begin{figure}
\plotone{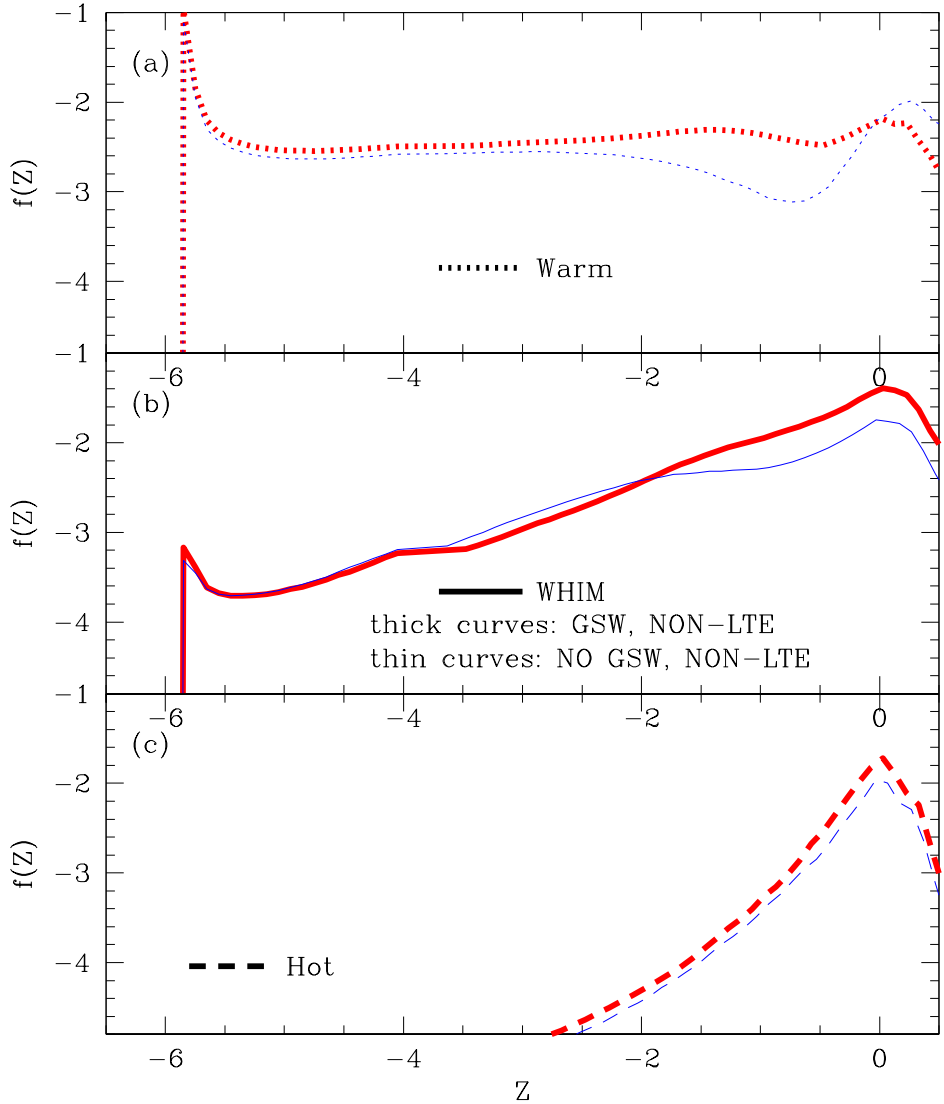}
\caption{
shows the differential 
mass fraction as a function of gas metallicity
for the three IGM components
for the simulation with
galactic superwinds (thick curves)
and that without galactic superwinds (thin curves).
Metal rich gas in the warm component near galaxies
has been shock heated to the WHIM state.
}
\label{f5}
\end{figure}

\begin{figure}
\plotone{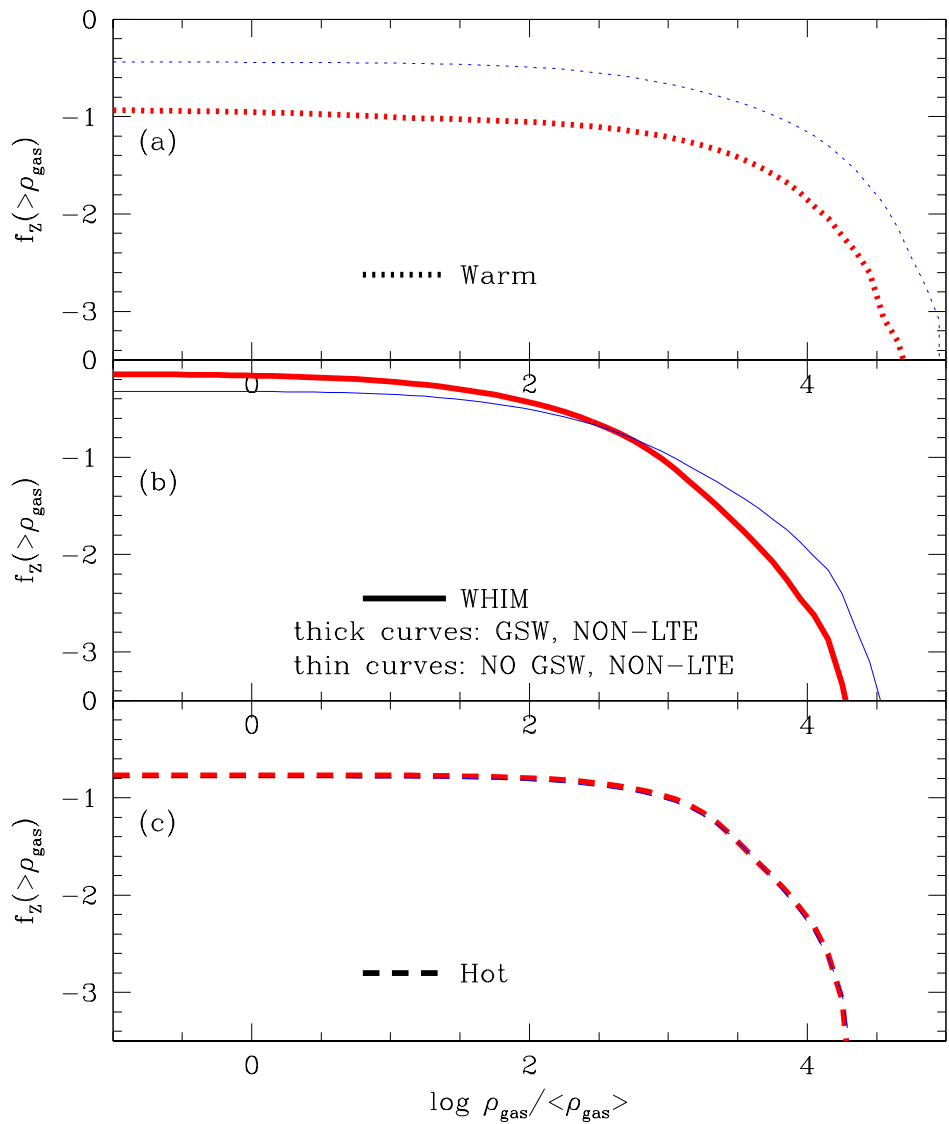}
\caption{
shows the cumulative metal
mass fraction as a function of gas density (in units of mean gas density)
for the three IGM components for the simulation with
galactic superwinds (thick curves)
and that without galactic superwinds (thin curves).
Note that the sum of $f_Z(>0)$ for the three component is unity in each
simulation.
}
\label{f5}
\end{figure}

\clearpage
\begin{figure}
\includegraphics[angle=0.0,scale=0.60]{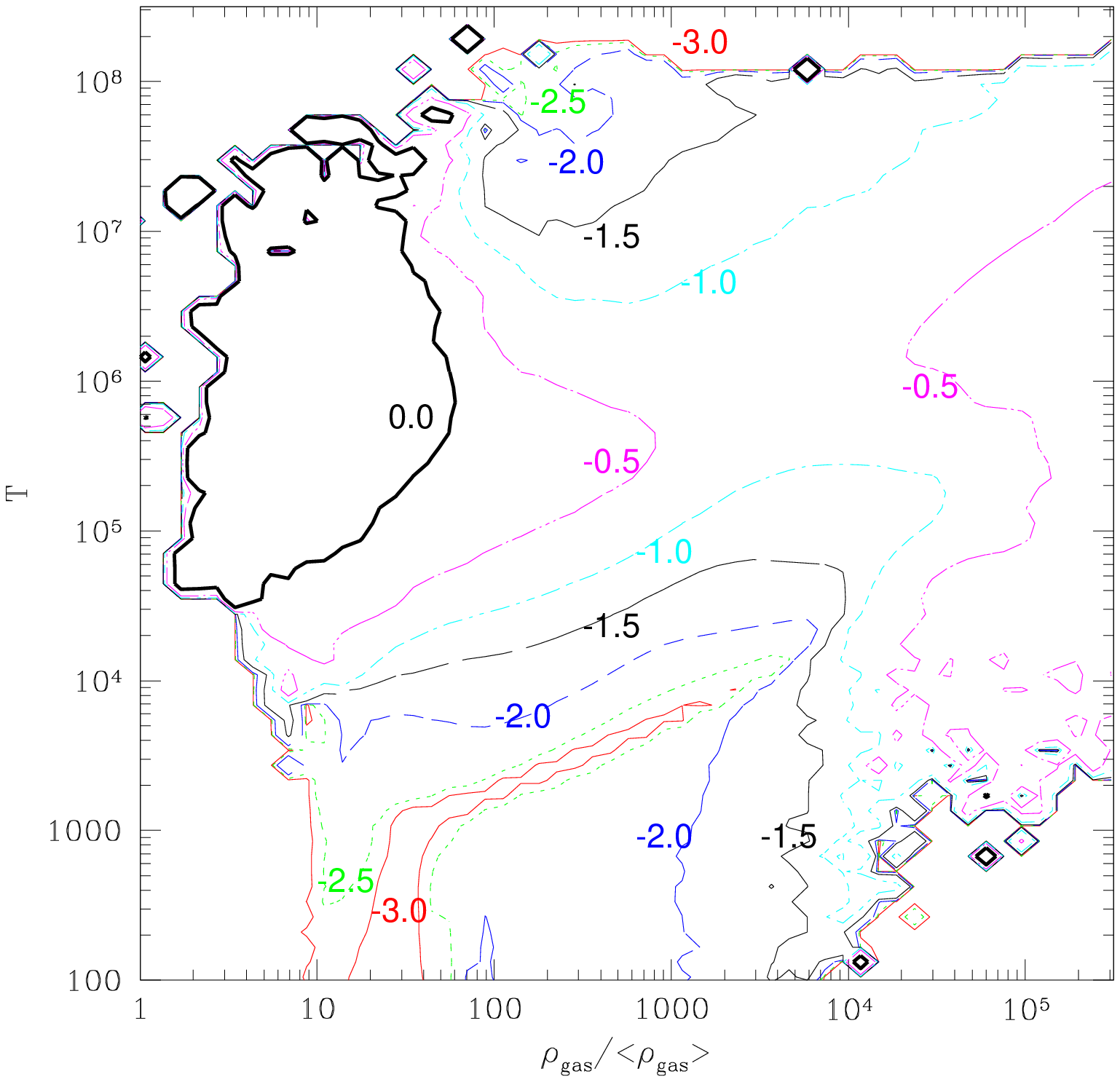}
\vskip -2.5cm
\includegraphics[angle=0.0,scale=0.60]{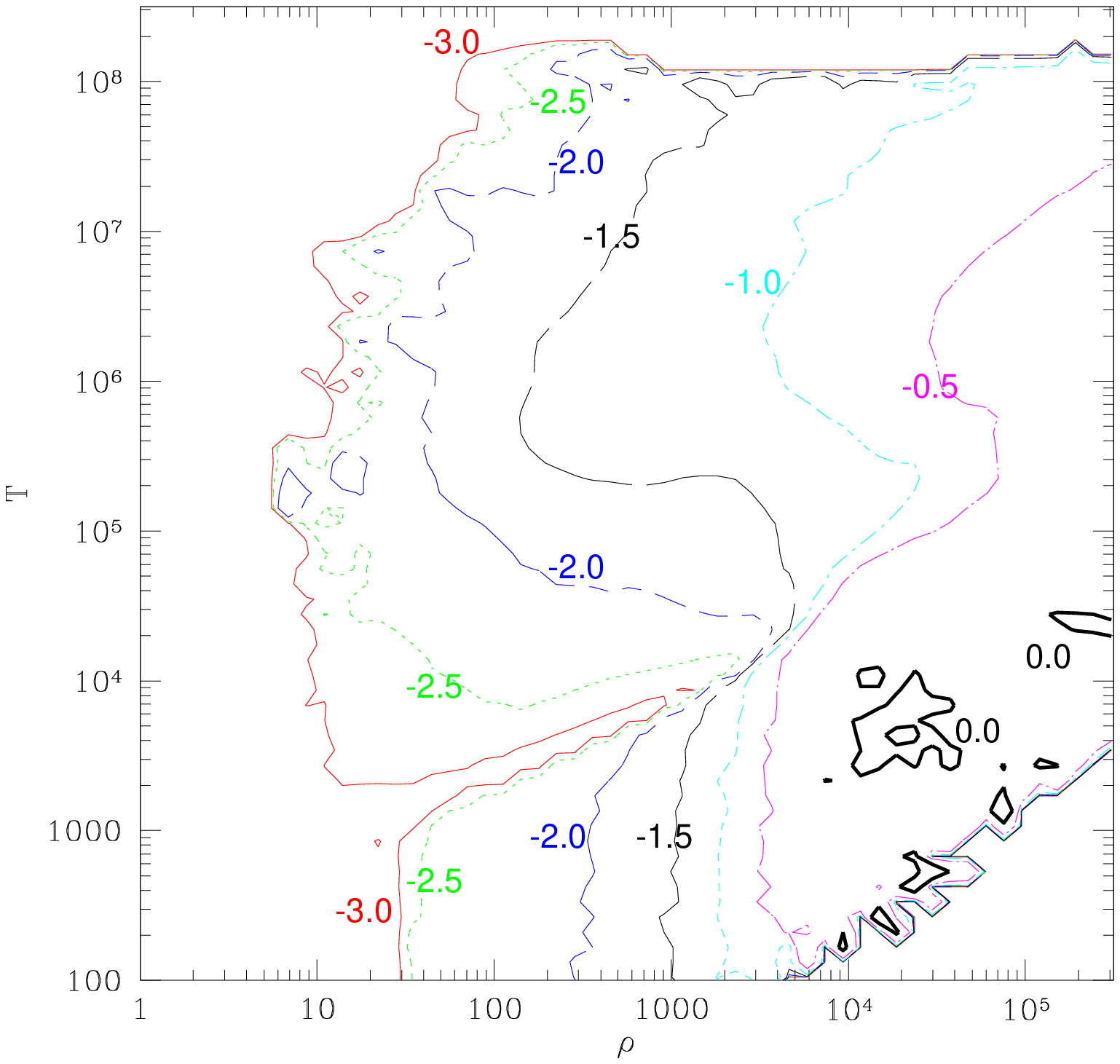}
\vskip -1.5cm
\caption{
shows the average metallicity of gas in the density-temperature plane
for the run with galactic superwinds (top panel)
and without (bottom panel).
The difference contours represent different metallicity levels,
labeled by numbers in solar units. 
Specifically, the thin solid, dotted, dashed, long-dashed,
dotted-dashed, dotted-long-dashed and thick solid contours
have metallicities of (-3.0, -2.5, -2.0, -1.5, -1.0, -0.5, 0.0) in solar units.
}
\label{f3}
\end{figure}

\end{document}